\documentclass[review]{elsarticle}
\usepackage[most]{tcolorbox}
\usepackage{lineno,hyperref}
\usepackage{verbatim}
\usepackage{placeins}
\usepackage{graphicx}
\usepackage{subcaption}
\modulolinenumbers[5]
\usepackage{amsmath}
\usepackage{amsmath}
\usepackage{amsthm}
\usepackage{amsfonts}
\usepackage{amssymb}
\usepackage{mathtools}

\journal{arXiv}

\biboptions{authoryear}

\usepackage[left=2cm,top=2cm,right=2cm,bottom=2cm,footskip=1cm]{geometry}
\usepackage{multirow}
\usepackage{diagbox}
\usepackage[ruled,linesnumbered,commentsnumbered]{algorithm2e}

\DeclarePairedDelimiter\abs{\lvert}{\rvert}%
\makeatletter
\let\oldabs\abs
\def\abs{\@ifstar{\oldabs}{\oldabs*}}
\makeatother
\begin{document}

\begin{frontmatter}

\title{Evolution of Novel Activation Functions in Neural Network Training with Applications to Classification of Exoplanets}

\author{Snehanshu Saha}
\address{PES University \\ snehanshusaha@pes.edu}
\author{Nithin Nagaraj}
\address{Consciousness Studies Programme \\ National Institute of Advanced Studies \\ nithin@nias.res.in}
\author{Archana Mathur}
\address{Indian Statistical Institute Bangalore \\  mathurarchana77@gmail.com}
\author{Rahul Yedida}
\address{North Carolina State University\\ yrahul3910@gmail.com}






\begin{abstract}
We present analytical exploration of novel activation functions as consequence of integration of several ideas leading to implementation and subsequent use in habitability classification of exoplanets. Neural networks, although a powerful engine in supervised methods, often require expensive tuning efforts for optimized performance. Habitability classes are hard to discriminate, especially when attributes used as hard markers of separation are removed from the data set.  The solution is approached from the point of investigating analytical properties of the proposed activation functions. The theory of ordinary differential equations and fixed point are exploited to justify the ''lack of tuning efforts'' to achieve optimal performance compared to traditional activation functions. Additionally, the relationship between the proposed activation functions and the more popular ones is established through extensive analytical and empirical evidence. Finally, the activation functions have been  implemented in plain vanilla feed-forward neural network to classify exoplanets.
\end{abstract}


\end{frontmatter}


\section{Introduction}
 Neural networks \citep{neural} or Artificial Neural network (ANN) are systems of interconnected units called ``Neurons'' organized in layers (loosely inspired by the biological brain). ANN is a framework for many different machine learning algorithms to process complex input data (text, audio, video etc.) in order to ``learn'' to perform classification/regression tasks by considering examples and without the aid of task-specific programming rules. As of today, ANNs are found to yield state-of-the-art performance in a variety of tasks such as speech recognition, computer vision, board games and medical diagnosis \citep{xiao2015finite,narayanan2004single} and classification of exoplanets~\citep{Insight}. 
 
Every neuron in ANNs is followed by an {\it activation function} which defines the output of that neuron given its inputs. It is an abstraction representing the rate of action potential firing in the neuron. In most cases, it is a binary non-linear function - the neuron either fires or not. The celebrated sigmoidal activation function enables a feed-forward network with a single hidden layer containing a finite number of neurons to approximate a wide variety of linear and non-linear functions~\citep{cybenko1989approximation}. Recent works have shown that activation functions allow neural networks to perform complex tasks, including gene expression analysis \citep{narayanan2004single} and solving nonlinear equations \citep{xiao2015finite}. The most popular activation functions are sigmoid, hyperbolic tangent (tanh) and ReLU (Rectified Linear Unit).

In this work, we explore novel activation functions as a consequence of integration of several ideas leading to implementation and subsequent use in habitability classification of exoplanets. The relationship between the proposed activation functions and the existing popular ones will be established through extensive analytical and empirical evidence. 

\subsection{Classification of Exoplanets}
Astronomers and philosophers have been intrigued by the fact that the Earth is the only habitable planet within the solar system. There has been scant evidence or explorations in the direction of finding planets outside the solar system which may harbor life. Essentially, the mission to find ''Earth 2.0'' gained momentum in recent times with NASA spearheading Kepler \citep{googlenasa} and other missions. Initial missions exploring Earth's neighbors, Mars and Venus, didn't yield promising results. However, NASA's missions in the past two decades led to discoveries of hundreds of exoplanets -- planets that are outside our solar system, also known as  extra-solar planets. The missions are carried out based on the inference that planets around stars are more frequent and the fact that actual number of planets far exceeds the number of stars in our galaxy. The mandate is to look for interesting samples from the pool of recently discovered exoplanet database \citep{phlref}. Additionally, finding ''Earth 2.0'' based on similarity metric \citep{googlenasa} \citep{CDHPF2016} is not the only approach. In fact, the approach has its own limitations as astronomers argue that a distant ``Earth-similar'' planet may not be habitable and could be just statistical mirage \citep{Keplercite}. Therefore, a classification approach should be used to vet habitability scores of exoplanets in order to ascertain the probability of distant planets harboring life. This begets the need for sophisticated pipelines. Goal of such techniques would be to quickly and efficiently classify exoplanets based on habitability classes. This is equivalent to a supervised classification problem.

Neural networks, although a powerful engine, often require expensive ``parameter-tuning" efforts. Classification of exoplanets is an intriguing problem and extremely hard, even for a neural network to solve, especially when clear markers for separation of habitability classes ( i.e. features such as surface temperature and features related to surface temperature) are removed from the feature set. To this end, a version of ANNs, replicated weight neural network (RWNN), in conjunction with sigmoid activation, has been applied to the task of classification of exoplanets \citep{Insight}. However, the results turned out to be less than satisfactory. Add to that, the effort in parameter tuning \citep{sbaf}, the outcome is far from desired when applied to pruned feature sets used in this paper (please refer to the Data section).

The process of discovery of exoplanets involves very careful analysis of stellar signals and is a tedious and complicated process, as outlined by Bains et.al. \citep{Bains2016}, . This is due to the smaller size of exoplanets compared to other types of stellar objects such as stars, galaxies, quasars, etc.  Radial velocity-based techniques and gravitational lensing are most popular methods to detect exoplanets. Given the rapid technological improvements and the accumulation of large amounts of data, it is pertinent to explore advanced methods of data analysis to rapidly classify planets into appropriate categories based on the physical characteristics.

The habitability problem has been tackled in different ways. Explicit Earth-similarity score computation \citep{CDHPF2016} based on parameters mass, radius, surface temperature and escape velocity developed into Cobb-Douglas Habitability Score helped identify candidates with similar scores to Earth. However, using Earth-similarity alone \citep{1804.11176} to address habitability is not sufficient unless model based evaluations \citep{1803.04644} are interpreted and equated with feature based classification \citep{Insight}. However, when we tested methods in \citep{Insight}, it was found that the methods didn't work well with pruned feature sets (features which clearly mark different habitability classes were removed). Therefore, new machine learning methods to classify exoplanets are a necessity.

To address this need for efficient classification of exoplanets, we embark on design of novel activation functions with sound mathematical foundations. We  shall justify the need to go beyond Sigmoid, hyperbolic tangent and ReLU activation functions and demonstrate the superior performance of the proposed activation functions in habitability classification of exoplanets. Such a fundamental approach to solving a classification problem has obvious merits and the activation functions that we propose could be readily applied to problems in other areas as well.

One of the key reasons to classify exoplanets is the limitation of finding out habitability candidates by Earth similarity alone, as proposed by several metric based evaluations, namely the Earth Similarity Index,  Biological Complexity Index \citep{BCI}, Planetary Habitability Index \citep{phlref} and Cobb-Douglas Habitability Score (CDHS)  \citep{CDHPF2016}. 
In \cite{Potential}, an advanced tree-based classifier, Gradient Boosted Decision Tree was used to classify Proxima b and other planets in the TRAPPIST-1 system. The accuracy was near-perfect, providing a baseline for other machine classifiers. However, that paper considered surface temperature as one of the attributes/features, making the classification task significantly easier than the proposed one in the current manuscript.


\section{Motivation and Contribution}
Observing exoplanets is no easy task. In order to draw a complete picture of any
exoplanet, observations from multiple missions are usually combined. For instance,
the method of radial velocity can give us the minimum mass of the planet and the
distance of the planet from its parent star, transit photometry can provide us with
information on the radius of a planet, and the method of gravitational lensing can
provide us the information regarding the mass. In contrast to this, stars are generally
easier to observe and profile through various methods of spectrometry and photometry in different ranges of wavelengths of the electromagnetic spectrum. Hence, by
the time a planet has been discovered around a star, we would possibly have fairly
rich information about the parent star, and this includes information such as the
luminosity, radius, temperature, etc. Therefore, the uniqueness of the approach is to classify
exoplanets based on smaller sets of observables as features of the exoplanet along with multiple features of the parent star. We elaborate each case of carefully selected features in the Experiments section. We note, most of the sub-sets of features created from PHL-EC do not contain surface temperature. In fact, the most interesting result of the paper might be that
some of the methods are unable to exactly reproduce the correct classification, given
the strict dependence of the classification on a single feature, surface temperature namely. Surface temperature is a hard marker when it comes to classifying exoplanets. Removing this feature enhances the complexity of classification many folds. Our manuscript tackles the challenge by adopting a foundational approach to activation functions.
\par However a reasonable question arises. When there exists activation functions in literature with evidence of producing good performance, is there a need to define a new activation function? If indeed the necessity is argued, can the new activation function be related to existing ones? We are motivated by these questions and painstakingly address these in subsequent sections 4-9 in the manuscript. We show the activation functions proposed in the paper, SBAF and A-ReLU are indeed generalizations of popular activations, Sigmoid and ReLU respectively. Moreover, the theory of Banach spaces and contraction mapping have been exploited to show that SBAF can be interpreted as a solution to first order differential equation. Further, the theory of fixed point and stability has been used to explain the amount of effort saved in tuning parameters of the activation units. In fact, this is one of the hallmarks of the paper where we intend to show that our activation functions implemented to tackle the classification problem are ``thrifty" in comparison to Sigmoid and ReLU. We demonstrate this by comparing system utilization metrics such as runtime, memory and CPU utilization. Additionally, we show that SBAF also satisfies the Universal Approximation Theorem and can be related to regression under uncertainty. We also demonstrate mathematically and otherwise that the approximation to ReLU, defined as A-ReLU is continuous and differentiable at "knee-point" and establish the fact that A-ReLU doesn't require much parameter tuning. Extensive experimentation shows conclusively, the insights gained from the mathematical theory are backed up by performance measures, beating Sigmoid and ReLU.

\par The remainder of the paper is organized as follows. A novel activation function to train an artificial neural network (ANN) is introduced. We discuss the theoretical nuances of such a function in section 5. We relate SBAF to binary logistic regression in section 6. The following section presnts a mathematical treatment of the evolution of SBAF from theory of differential equations. Next section introduces A-ReLU as an approximation exercise. We follow up with network architecture in the the next section, detailing the back propagation mechanism  paving the foundation for ANN based classification of exoplanets. Consequently, the following section presents results on all data sets with performance comparison including system utilization. We conclude by discussing the efficacy of the proposed method.



\section{Saha-Bora Activation Function (SBAF)}
We shall introduce SBAF neuron to address the issues elicited in the previous sections. We begin with a brief description of mathematical concepts and definitions will be used throughout the remainder of the manuscript.
\begin{tcolorbox}[colback=red!5!white,colframe=red!75!black]
\textit{Definition of key terms:}
\begin{itemize}
\item Returns to Scale: For the objective function, $y=kx^\alpha(1-x)^\beta, k>0, 0<\alpha<1, 0<\beta<1 $ feasible solution that maximizes the objective function exists, called an optimal solution under the constraints {\bf Returns to scale}. When $\alpha+\beta>1$, it is called increasing returns to scale (IRS) and $\alpha+\beta<1$ is known as decreasing returns to scale (DRS). $\alpha+\beta=1$ is known as constant returns to scale and ensures proportional output, $y$ to inputs $x, 1-x$. $y$ is used to define production functions in Economics and the form is inspirational to designing the activation functions proposed in the manuscript. Note, we set $\beta=1-\alpha$ in the activation function formulation ensuring CRS and therefore existence of global optima.
    \item Absolute error is the magnitude of the difference between the exact value and the approximation.
    \item Relative error is the absolute error divided by the magnitude of the exact value.
    \item Banach Space: A complete normed vector space i.e. the Cauchy sequences have limits. A Banach space is a vector space.
    \item Contraction Mapping: Let $(X,d)$  be a Banach space equipped with a distance function d. There exists a transformation, T such that $T:X\rightarrow X$ is a contraction, if there is a guaranteed $q<1,q~ \text{may be zero}$ which squashes the distance between successive transformations (maps). In other words, $d(T(x), T(y))< qd(x,y)\hspace{0.2mm}~~ \forall x,y \in X$. $q$ is called Lipschitz constant and determines speed of convergence of iterative methods.
    \item Fixed point: A fixed point $x_*$ of a function $f: \mathbb{R} \rightarrow \mathbb{R}$ is a value that is unchanged by repeated applications of the function, i.e., $f(x_*) = x_*$. Banach space endowed with a contraction mapping admits of a unique fixed point. Note, for any transformation on Turing Machines, there will always exist Turing Machines unchanged by the transformation.
    \item Stability: Consider a continuously differentiable function $f: \mathbb{R} \rightarrow \mathbb{R}$ with a fixed point $x_*$, $f(x_*)=x_*$. The fixed point $x_*$ is defined to be {\it stable} if $f'(x_*) < 1$. If $f'(x_*) > 1$, then $x_*$ is defined as {\it unstable}. 
    \item First Return Map:  Consider an iterative map on the set $S$, $f: S \rightarrow S$. Starting from an initial value $x_0 \in S$, we can iterate the map $f$ to yield a trajectory: $x_1 = f(x_0)$, $x_2 = f(x_1)$ and so on.  The first-return map is a plot of $x_{n+1}~vs.~x_n$ for integer $n>0$ .  
    \item Contour Plot: A contour plot is a 2-D representation of a 3-D surface. An alternative to surface plot, it provides valuable insights to changes in $y$ as input variables $x \& 1-x$ change.
\end{itemize}
\end{tcolorbox}

This section defines SBAF, computes its derivative and focuses on estimation of parameters used in the function. We compute the derivative of SBAF for two reasons: to use the derivative in back-propagation and to show that SBAF possesses optima instead of saddle point (note, sigmoid has saddle point). SBAF,  defined with parameters $k, \alpha$ (these will be estimated in subsequent sections and no effort is spent on tuning these parameters while training) and input $x$ (Data in the input layer), produces output $y$ as follows:
\begin{equation}
\begin{split}
y &= \frac{1}{1 + kx^{\alpha}(1-x)^{1-\alpha}}; \\
\Rightarrow \textrm{ln}y &= \textrm{ln}1 - \textrm{ln}(1 + kx^{\alpha}(1-x)^{1-\alpha})\\
 &= - \textrm{ln}(1 + kx^{\alpha}(1-x)^{1-\alpha})\\
\Rightarrow \frac{1}{y}\frac{dy}{dx} &= - \frac{1}{(1 + kx^{\alpha}(1-x)^{1-\alpha})} \cdot \Big[k\alpha x^{\alpha -1} (1-x)^{1-\alpha} - kx^{\alpha}(1-\alpha)(1-x)^{1 - \alpha -1} \Big]\\
&= - \frac{k}{(1 + kx^{\alpha}(1-x)^{1-\alpha})} \cdot \Big[\alpha x^{\alpha -1} (1-x)^{1-\alpha} - (1-\alpha)x^{\alpha}(1-x)^{-\alpha} \Big]\\
\Rightarrow \frac{dy}{dx} &= -y^2 \Bigg[ \frac{\alpha}{x} - (1 - \alpha)\frac{1}{1-x} \Bigg]kx^{\alpha}(1-x)^{1-\alpha}\\
&=- y^2 \Bigg[\frac{\alpha (1-x) - (1-\alpha)x}{x(1-x)} \Bigg]kx^{\alpha}(1-x)^{1-\alpha}\\
&= -y^{2} \Bigg[\frac{\alpha - x}{x(1-x)} \Bigg]kx^{\alpha}(1-x)^{1-\alpha}\\
\end{split}
\label{eq:acti_func_1}
\end{equation}

From the definition of the function, we have:

\begin{equation}
\begin{split}
&y = \frac{1}{1 + kx^{\alpha}(1-x)^{1-\alpha}}\\
\Rightarrow& kx^{\alpha}(1-x)^{1-\alpha} = \frac{1-y}{y}
\end{split}
\label{eq:acti_func_2}
\end{equation}

Substituting Equation \ref{eq:acti_func_2} in \ref{eq:acti_func_1},

\begin{equation}
\begin{split}
\frac{dy}{dx} &= -y^{2} \cdot \frac{\alpha-x}{x(1-x)} \cdot \frac{1-y}{y}\\
&= \frac{y(1-y)}{x(1-x)}\cdot(x-\alpha)
\end{split}
\label{eq:acti_func_3}
\end{equation}

\begin{figure}[htbp!]
\begin{center}
\includegraphics[width=0.6\columnwidth]{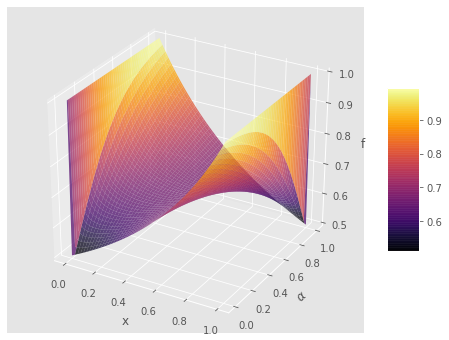} 
\caption{Surface Plot of SBAF: The x-axis shows $x$, and the $y$ axis shows $\alpha$, both varied from 0 to 1. $k$ is fixed to 1.}
\label{fig:sbaf:1}
\end{center}
\end{figure}

\begin{figure}[htbp!]
\begin{center}
\includegraphics[width=0.5\columnwidth]{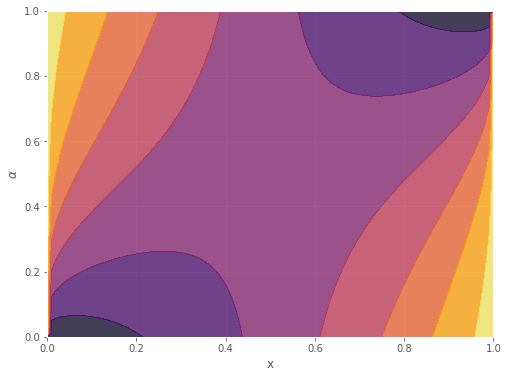} 
\caption{Contour plot for SBAF, with $x$ and $\alpha$ varied from 0 to 1. $k$ is fixed to 1.}
\label{fig:sbaf:2}
\end{center}
\end{figure}

Figures \ref{fig:sbaf:1} and \ref{fig:sbaf:2} show various plots of the activation function. Figure \ref{fig:sbaf:1} shows a surface plot by varying $x$ and $\alpha$. Figure \ref{fig:sbaf:2} shows the contour plot. In both plots, we fixed $k=1$. Note that the contour plot reveals a symmetry about $x = \alpha$. Further, as discussed in Section \ref{sec:approximating}, the approximation of other activation functions using SBAF can be quite easily seen in the contour plot, by looking at the horizontal lines corresponding to $\alpha=0$ and $\alpha=1$.

We note, the motivation of SBAF is derived from using $kx^{\alpha}(1-x)^{1-\alpha}$ as discriminating function to optimize the width of the two separating hyperplanes in an SVM-like formulation. This is equivalent to the CDHS formulation when CD-HPF \citep{CDHPF2016} is written as $y = kx^{\alpha}(1-x)^{\beta}$ where $\alpha+\beta=1, 0 \leq \alpha \leq 1, 0 \leq \beta \leq 1$, and $k$ is suitably assumed to be $1$ (CRS condition). The representation ensures global maxima (maximum width of the separating hyperplanes) under such constraints \citep{CDHPF2016}. Please also note, when $\beta=1-\alpha$, CRS condition \citep{Potential} is satisfied automatically and we obtain the form of the activation, SBAF.

Our activation function has an optima. From the visualization of the function below, we observe less flattening of the function, tackling the ``vanishing gradient" problem. Moreover, since $0 \leq \alpha \leq 1, 0 \leq x \leq 1, 0 \leq 1-x \leq 1$, we can approximate, $kx^{\alpha}(1-x)^{1-\alpha}$ to a first order polynomial. This helps us circumvent expensive floating point operations without compromising the precision.

\begin{figure}[htbp!]
\begin{center}
\includegraphics[width=0.5\columnwidth]{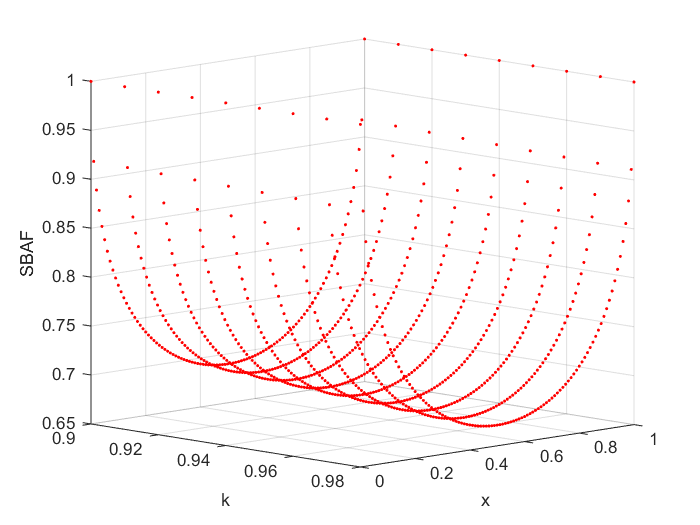} 
    \caption{Plot of SBAF function: it shows a minima for all values of $k (0.90-0.98)$ at $x=0.5$. Empirical test for stability confirms stable fixed point at $k=0.98, \alpha=0.5$ (Note, mimima of SBAF is obtained at $x=\alpha$ and thus we set $\alpha=0.5$). These are the $k, \alpha$ values we used in SBAF to train the classifier.}
\label{fig:sbaf2}
\end{center}
\end{figure}

\subsection{Existence of Optima: Second order Differentiation of SBAF for Neural Network}

From Equation \ref{eq:acti_func_3} ,

\begin{equation}
\begin{split}
\frac{dy}{dx} &=\frac{y(1-y)}{x(1-x)}\cdot(x-\alpha)
\end{split}
\nonumber
\end{equation}
It's easy to see,

\begin{equation}
\begin{split}
\Rightarrow \frac{d^{2}y}{dx^{2}} &=- \frac{x(1-x)(\alpha-x)\cdot(1-2y)\cdot\Big[\frac{y(1-y)}{x(1-x)}\cdot(\alpha-x)\Big] + x(1-x)\cdot y(y-1) + y(y-1)\cdot(\alpha-x)\cdot(1-2x)}{(x(1-x))^{2}}\\
&= -\frac{y(y-1)[x(1-x)+(\alpha-x)\cdot (1-2x) + (\alpha-x)^{2}\cdot (1-2y)]}{(x(1-x))^{2}} \\
\end{split}
\nonumber
\end{equation}
when $\alpha = x$, 
\begin{equation}
\begin{split}
\Rightarrow \frac{d^{2}y}{dx^{2}} &= - \frac{x(1-x)\cdot y(y-1)}{(x(1-x))^{2}} \\
&= \frac{y(1-y)}{x(1-x)} \\
\end{split}
\nonumber
\end{equation}

Clearly, the first derivative vanishes when $\alpha=x$, and the derivative is positive when $\alpha < x$ and is negative when $\alpha > x$ (implying range of values for $\alpha$ so that the function becomes increasing or decreasing). We need to determine the sign of the second derivative when $\alpha=x$ to ascertain the condition of optima (corresponding to minima of gradient descent training ensuring optimal discrimination between habitability classes). Assuming $ 0< x <1$, the condition of optimality, $0 \leq \alpha \leq 1$, $y$ by construction lies between $(0,1)$. Hence, $\frac{d^{2}y}{dx^{2}} > 0$ ensuring optima of $y$.
\subsection{Saddle points of sigmoid activation and a comparative note with SBAF}
We note briefly that as opposed to the sigmoid activation function, SBAF has local minima and maxima. For example, for $k=0.91, \alpha=0.5$, a local minima occurs at $x=\alpha$. On the other hand, the sigmoid function has neither local minima nor maxima; it is easy to show that it only has a saddle point at $x=0$: 

\begin{proof}
Note that if $f(x)$ denotes the sigmoid function, then $f^\prime(x) = f(x)(1-f(x)); f^{\prime\prime}(x) = f(x)(1-f(x))(1-2f(x))$. Then, $f^\prime(x)=0$ implies that $f(x) \in \{0, 1\}$. However, for both these values, the second derivative is 0.
\end{proof}

SBAF, however, has a critical point at $x=\alpha$.

\subsection{Universal Approximation Theorem for SBAF}
It is well known that a feed-forward network with a single hidden layer containing a finite number of neurons satisfies the universal approximation theorem. This is important since it guarantees that simple neural networks can represent a wide variety of interesting linear/non-linear functions (with appropriately chosen parameters). \cite{cybenko1989approximation} proved one of the first versions of this theorem for sigmoid activation functions. We show that SBAF is also sigmoidal and hence satisfies the Universal Approximation Theorem. 

Following \cite{cybenko1989approximation}, we shall define the following:
\begin{itemize}
\item $I_n$: $n$-dimensional unit cube, $[0,1]^n$.
\item $C(I_n)$: Space of continuous functions defined on $I_n$.
\item $M(I_n)$: The space of finite, signed regular Borel measures on $I_n$.
\item $\sigma(t)$: A univariate function is defined as being \emph{sigmoidal} if
\begin{eqnarray*}
\sigma(t) & \rightarrow 1  \text{ as } t \rightarrow +\infty,\\
\sigma(t) & \rightarrow 0  \text{ as } t \rightarrow -\infty.
\end{eqnarray*}
\item $\sigma$ is defined to be \emph{discriminatory} if for a measure $\mu \in M(I_n)$, 
\begin{equation*}
\int_{I_n} \sigma(y^Tx+\theta) ~~d\mu(x) = 0
\end{equation*}
$\forall y \in \mathbb{R}^n$ and $\theta \in \mathbb{R}$ implies that $\mu=0$.
\end{itemize}

\noindent \emph{Approximation Theorem (Cybenko, 1989): Let $\sigma$ be any continuous discriminatory function. Then finite sums of the form}
\begin{equation*}
G(x) = \sum_{j=1}^{N} \alpha_i \sigma(y_j^Tx + \theta_j)
\end{equation*}
\emph{are dense in $C(I_n)$. In other words, given any $f \in C(I_n)$ and $\epsilon >0$, there is a sum, $G(x)$, of the above form, for which}
\begin{equation*}
|G(x) - f(x)| < \epsilon ~~ \forall ~~ x\in I_n.
\end{equation*}

\begin{proof}
Please refer to Cybenko (1989).
\end{proof}
Also, it should be noted that by Lemma 1 of ~\cite{cybenko1989approximation}, \emph{any continuous sigmoidal function is discriminatory.} We can thus prove:\\
{\noindent} {\it Universal Approximation Theorem for SBAF:~} The proposed function $SBAF_{k,\alpha}(x)$ satisfies the universal approximation theorem.

\begin{proof}
We observe that $SBAF_{k,\alpha}(x)$ (for the choices $k=1$ and $\alpha=0$, see Section \ref{sec:approximating}) is a continuous sigmoidal function and hence discriminatory. All the conditions of the approximation theorem are met.
\end{proof}
Thus SBAF can be used with a feed-forward network with a single hidden layer containing a finite number of neurons to approximate a wide variety of linear and non-linear functions.

\subsection{SBAF is not a probability density function}
\begin{proof}
Let us compute the integral below:
$\int_{I}\frac{1}{1 + kx^{\alpha}(1-x)^{1-\alpha}}$ where I is the interval containing ($-\infty, +\infty$). We know from earlier calculations, that
\begin{equation}
\begin{split}
y &= \frac{1}{1 + kx^{\alpha}(1-x)^{1-\alpha}}; 
\end{split}
\end{equation}
\& 
\begin{equation}
\begin{split}
\frac{dy}{dx} &=\frac{y(1-y)}{x(1-x)}\cdot(x-\alpha); 
\end{split}
\end{equation}
Using the above in the integral and readjusting the limits of integration, we observe that $\int_{0,0}\frac{x(1-x)}{y(1-y) \cdot(x-\alpha)}dy \rightarrow 0$. Note, $0\leq f(x) \leq 1, \forall x \in (0,1).$ Therefore the integral of $f(x)$ between 0 and 1 would also be less than or equal to 1. Equality will hold iff $f(x)=1$, but this is not possible for any $x \in (0,1)$.
\end{proof} 

SBAF is sampled from a PDF. SBAF is not a PDF and we infer that it may oscillate. The next section contains a more pertinent insight, in clear agreement with the analysis, relating SBAF to regression under uncertainty.


\section{Relating SBAF to Binary Logistic Regression: Regression under Uncertainty}
Binary logistic regression builds a model to characterize the probability of $Y_i$ (observed value of the binary response variable $Y$) given the values of the explanatory variable $x_i$ in the following manner:
\begin{equation}
\log\left(\frac{\pi_i}{1-\pi_i}\right) =  \alpha_0  + \alpha_1 x_i,
\label{eqn:eqn_logi1}
\end{equation}
where $\pi_i = P(Y_i=1|x_i=x)$ denotes the probability that the response variable $Y_i=1$ given the value $x_i=x$, and $i$ stands for the index of the observed sample.  The LHS of equation~\ref{eqn:eqn_logi1} is known as \emph{logit}($\cdot$) or \emph{log-odds}. The above formulation leads to the celebrated sigmoid activation function:
\begin{equation}
\pi_i = \frac{\exp(\alpha_0  + \alpha_1 x_i)}{1+ \exp(\alpha_0  + \alpha_1 x_i)},
\end{equation}
where the regression co-efficients $\alpha_0$ and $\alpha_1$ are to be determined.

Instead, if we formulate the logit as: 
\begin{equation}
\log\left(\frac{\pi_i}{1-\pi_i}\right) =  - \alpha \log(x_i) - (1-\alpha) \log(1-x_i) - \log(k),
\label{eqn:eqn_logi2}
\end{equation}
where $0 < x_i < 1$, constants $k>0$ and $0 \leq \alpha \leq 1$, then it leads to our proposed activation function:
\begin{equation}
\pi_i = \frac{1}{1 + kx_i^{\alpha}(1-x_i)^{1-\alpha}}
\label{eq:maxent:1}
\end{equation}

Equation~\ref{eqn:eqn_logi2} can be understood as follows. Firstly, think of $\{ x_i \}$ as probabilities of the observed explanatory variable $X$ instead of the value itself. If $X$ is a binary discrete random variable,  then the quantities $-\log(x_i)$ and $-\log(1-x_i)$ would be the {\it self-information} of these two outcomes (measured in bits if the base of the logarithm is 2). The convex combination of these two self-information quantities is like an {\it average information quantity (Shannon entropy)}. In fact, the RHS is upper bounded by Shannon entropy (if $\alpha = x_i$ then RHS is the entropy of $\{ x_i\}$).

\begin{proof}
Maximizing \eqref{eq:maxent:1} is equivalent to minimizing its inverse. We ignore the constant term 1, and assume that $k$ is fixed. Then, the minima occurs where the derivative is 0:

\begin{align*}
    \frac{\mathrm{d}}{\mathrm{d}x} \left( kx^\alpha (1-x)^{1-\alpha} \right) &= k\alpha x^{\alpha-1}(1-x)^{1-\alpha} - k(1-\alpha)x^\alpha (1-x)^{-\alpha} \\
    &= kx^{\alpha-1}(1-x)^{-\alpha}\left( \alpha(1-x) - (1-\alpha)x \right) \\
    &= kx^{\alpha-1}(1-x)^{-\alpha} (\alpha - x)
\end{align*}
Clearly, the minima of this, and therefore the maxima of \eqref{eq:maxent:1}, occurs when $\alpha = x_i$.
\end{proof}

The quantity $-\log(k)$ can be thought of as a bias term or the information content that is universally available (if we think of the RHS as average uncertainty instead of average information, then this quantity would be the irreducible uncertainty that is present in the environment, such as noise). Thus, under this scenario, the log-odds of classifying the binary response variable $Y$ as 1 is a function of the average self-information of the observed explanatory variable. 

This means that if the probability of observed explanatory variable was very close to zero, then $x_i$ would be close to zero and the RHS would also be close to zero giving the log-odds ratio (LHS) also close to zero. If the probability of the observed explanatory variable was close to 0.5, then RHS would be very high and the log-odds ratio (RHS) would be close to 1.  Now, as the probability increases towards 1, the RHS starts to reduce and log-odds ratio (LHS) also starts to reduce. This means that we trust the observed variable only if it has probability between 0 and 0.5.  We do not trust high values (above 0.5).  

Another way to interpret this is that as the uncertainty increases, then the neuron fires (activates). The neuron in this case continuously increases the magnitude of its response as the uncertainty increases. The term $-\log(k)$ is the ambient noise term, and the neuron can fire if this is high enough (greater than some preset threshold $T$). 
The task of regression here is to determine $\alpha$ and $k$ such that it models the binary response variable $Y$ as a function of the uncertainty of the observed variables (along with noise in the environment). 


\section{Functional Properties of SBAF}
\subsection{Existence of a fixed point}

Let's consider the first order differential equation
\begin{equation}
\frac{\mathrm{d}y}{\mathrm{d}x} = \frac{y(1-y)}{x(1-x)}\cdot(x-\alpha)
\end{equation}
which is a representation of the standard form:
\begin{equation}
\frac{\mathrm{d}y}{\mathrm{d}x} = f(x,y)
\end{equation}
In order to prove the existence of a fixed point, we need to show that $f(x,y)$ is a Lipschitz continuous function. On differentiating y w.r.t x we obtain y as:
\begin{equation}
f(x,y) = \frac{\mathrm{d}y}{\mathrm{d}x} = \frac{y(1-y)}{x(1-x)}*(x-\alpha) 
\end{equation}
When $0<x<1$, we observe that $y<1$. Moreover, when $x \rightarrow 0$, $y \rightarrow 1$ and when $x \rightarrow 1$, $y \rightarrow 1$. In our case, $f$ is a continuous and differentiable function over the interval [0,1]. This implies f is bounded in (0,1). This further implies that the function follows the mean value theorem. That is, for some $c \in (0, 1)$,
\begin{equation*}
f^\prime(c) =  \frac{f(b)-f(a)}{b-a}
\end{equation*}
On differentiating $f$ w.r.t $x$ we obtain the following equation
\begin{equation}
f^\prime(x,y) = \frac{\mathrm{d}y}{\mathrm{d}x} = \frac{y(1-y)}{x(1-x)}    
\end{equation}

Since we already know the representation of $y$ in terms of $x$ from \eqref{eq:maxent:1}, we can substitute the value of $y$ and we obtain the following representation for $f$: 
\begin{equation*}
f^\prime(x) = \frac{kx^\alpha(1-x)^{(1-\alpha)}}{(1+kx^\alpha(1-x)^{(1-\alpha)})^2} *\frac{1}{x(1-x)}
\end{equation*}

Clearly, the right hand side is bounded between $(0,1)$ by some positive constant $K$. Using the above constraint in the mean value theorem specified, we obtain the following:
\begin{align*}
f^\prime(c) &= \frac{f(b)-f(a)}{-a}<=K\\    
f(b)-f(a) &\leq K(b-a)\\
|f(b)-f(a)| &\leq K
\end{align*}
Since $a=0$ and $b=1$, the above equation is of the form
\begin{align*}
|f(b)-f(a)| \leq K|(b-a)|    
\end{align*}
Therefore, $f$ is a Lipschitz continuous and by Picard's theorem, the differential equation has a fixed point and a unique solution in the form of the activation function, to be used in neural networks for classification.

\subsection{Lipschitz continuity, contraction map and unique fixed point}
In this section, we show that the fixed point is unique in the interval, thus ensuring that the solution to the DE, SBAF, is unique in that interval. We exploit the Banach contraction mapping theorem to achieve this goal. We consider the following DE
\begin{align}
f(x,y) &= \frac{\mathrm{d}y}{\mathrm{d}x} = \frac{y(1-y)}{x(1-x)}\cdot (x-\alpha) \\
y(x_0) &= y_0 \nonumber
\end{align}
Assume  $f(x,y)$ to be continuous on $D = (x_0 - \delta, x_0 + \delta), (y_0-b, y_0+b)$. Then $\exists$ a solution in $D$. This gives rise to the solution as the activation function, $y$. Furthermore, if $f(x,y)$ is Lipschitz continuous in D, or in a region smaller than D, the solution thus found is unique.
 
\begin{proof}
(a) We first prove that $f(x, y)$ is Lipschitz continuous. We write $y^\prime= \frac{\mathrm{d}y}{\mathrm{d}x}= f(x,y)= \frac{y(1-y)}{x(1-x)}*(x-\alpha)$.

Now,
\begin{align*}
\abs{f(x,y_1)-f(x,y_2)} &= \abs{\frac{x-\alpha}{x(1-x)}\left( y_1(1-y_1)-y_2(1-y_2) \right)}
\end{align*}

Within $(0,1)$, $\abs{\frac{x-\alpha}{x(1-x)}}$ is bounded. The expression blows up at $x=0, 1$. Therefore,
\begin{align*}
\abs{f(x,y_1)-f(x,y_2)} &\leq k\abs{y_1-y_1^2 -y_2+y_2^2} \\
\abs{f(x,y_1)-f(x,y_2)} &\leq k\abs{y_1 - y_2 - (y_1-y_2)(y_1+y_2)} \\
\abs{f(x,y_1)-f(x,y_2)} &\leq k\abs{(y_1 - y_2)(1-y_1-y_2)} \\
\end{align*}
By construction, $y_1$ and $y_2$ are bounded by some positive constant $\xi \in (0,1)$. Therefore, the sum is bounded by $2\xi$ and $\abs{1-(y_1+y_2)}\leq 1 \Rightarrow \abs{f(x,y_1)-f(x,y_2)} \leq k\abs{y_1-y_2}$

This implies $f(x,y)$ is Lipschitz continuous in $y$. We may now proceed to establish the existence of a unique fixed point i.e unique solution (activation function) to the differential equation above.

(b) Let T be a contraction mapping, i.e. $T:X \rightarrow X$, where $X$ is a complete metric space. Then $T$ has a unique fixed point in $X$ \citep{metric}. Moreover, let $x_0 \in X$, we define ${x_k}$ by setting an iterative map, $x_{k+1}=T(x_k)$. Let us fix $d_0 = d(x_0,x_1)$. Then, by Lipschitz continuity

\begin{align*}
d(x_k,x_{k+1}) &= d(T(x_{k-1}),T(x_{k})) \leq \alpha_1d(x_{k-1},x_{k}) \\
d(x_k,x_{k+1}) &\leq \alpha_1d(T(x_{k-2}),T(x_{k-1})) \\
d(x_k,x_{k+1}) &\leq \alpha_1^2d((x_{k-2}),x_{k-1})) \\
& \vdots \\
d(x_k,x_{k+1}) &\leq \alpha_1^kd(x_{0},x_{1}) =\alpha_1^kd_0  \\
\end{align*}

Clearly, ${x_k}$ is a Cauchy sequence. Since $X$ is a complete metric space, ${x_k}\rightarrow x \in X$. Thus,

\begin{align*}
d(T(x_k),T(x)) \leq \alpha d(x_k,x) \rightarrow 0
\end{align*}

Thus, $T(x_k) \rightarrow T(x)$. But $T(x_k) = x_{k+1}$. Therefore, $T(x) \rightarrow x$, i.e., $T(x) =x$. Suppose $T(y) = y$ for $y \neq x$.

\begin{align*}
d(x,y) = d(T(x),T(y)) &\leq \alpha d(x,y)\\
d(x,y) &\leq \alpha d(x,y)
\end{align*}

This is possible only if $d(x,y) = 0 \Rightarrow x=y$. This implies that the fixed point is unique.
\end{proof}

We use the following lemma to complete the missing piece. $f$ is continuous and Lipschitz w.r.t $y$ on the defined domain. Then, there exists a unique fixed point of $T$ in $X$. The function $y$ (activation function) is the unique solution. We establish the fact that the activation function is unique in the interval (0,1).

\subsection{Computation of the fixed point}
Now that the existence of of the fixed point is established, we proceed to find it in the following manner.
Let us consider the representation of y given by (1). By definition of fixed point we have $T(x)= y(x) = x$ at the fixed point, i.e.

\begin{align}
x = \frac{1}{1+kx^\alpha(1-x)^{1-\alpha}} &= T(x) \nonumber\\
x[1+kx^\alpha(1-x)^{1-\alpha}] &= 1 \nonumber\\
x+kx^{\alpha+1}(1-x)^{1-\alpha} &= 1 \nonumber\\
kx^{\alpha+1}(1-x)^{1-\alpha} &= 1-x \nonumber\\
kx^{\alpha+1}(1-x)^{-\alpha} &= 1 \label{eq:comp:1}
\end{align}

Assume $k =1$. Hence, \eqref{eq:comp:1} becomes

\[
x^{\alpha+1}(1-x)^{-\alpha} = 1
\]

On applying log on both sides we obtain

\begin{align*}
\log \left(x^{\alpha+1}(1-x)^{-\alpha} \right) &=0\\
\log x^{\alpha+1}+\log (1-x)^{-\alpha}&=0\\
(\alpha+1)\log x -\alpha \log (1-x)&=0\\
\alpha \log x + \log x -\alpha \log (1-x)&=0\\
\alpha (\log x-\log (1-x)) + \log x&=0\\
\log \left(\frac{1-x}{x} \right)^\alpha &=\log x\\
\end{align*}

Thus, $\alpha =\frac{\log x}{\log \frac{1-x}{x}}$.

\subsection{Visual Analysis of the fixed point }
We use the formula from the above computation and visually represent the activation function, SBAF, the solution to the differential equation mentioned above. We observe that, for K=0.9 onward, we obtain a stable fixed point. SBAF used for training the neural net is able to classify PHL-EC data with remarkable accuracy when K is very close to 1. Existence of a stable fixed point thus is a measure of classification efficacy, in this case.

We explore the non-linear dynamics of SBAF. At a fixed point, as noted above, we have:
\begin{align*}
kx^{\alpha+1}(1-x)^{-\alpha}=1.
\end{align*}
In the above expression for the fixed point, for all real values of $\alpha$ and when $0 < x < 1$, the left-hand side is always negative if $k<0$. Thus, there can't be any fixed point for $k<0$ when $0 < x < 1$.  When $k>0$, it is possible to have a fixed point $x^*$. Below we plot the first return maps for the SBAF for a few cases with  fixed $\alpha = 0.5$ and for different values of $k$. We plot only the real values of $x$ (since we have complex dynamics as $x$ can become a complex number).

\begin{figure}[htbp!]
\begin{center}
\includegraphics[width=0.40\columnwidth]{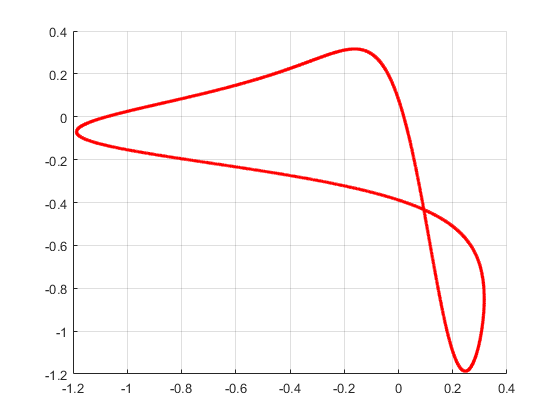}
\includegraphics[width=0.40\columnwidth]{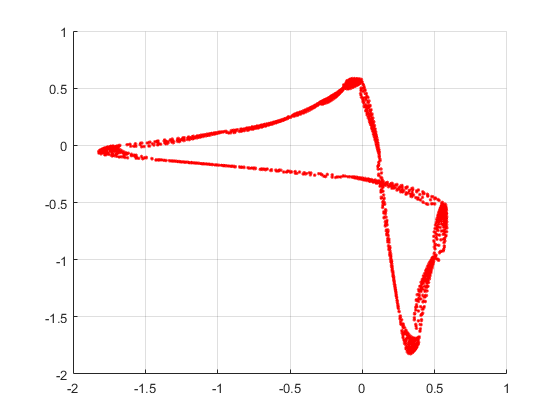}\\
\includegraphics[width=0.40\columnwidth]{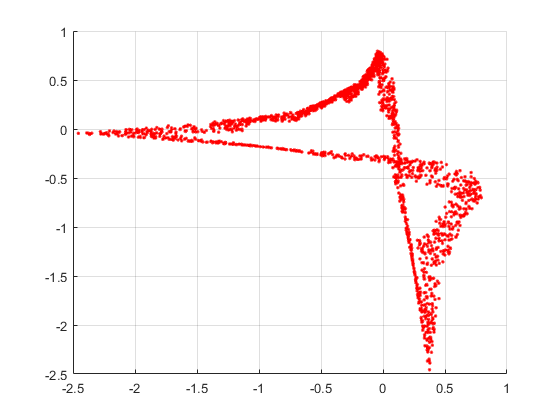}
\includegraphics[width=0.40\columnwidth]{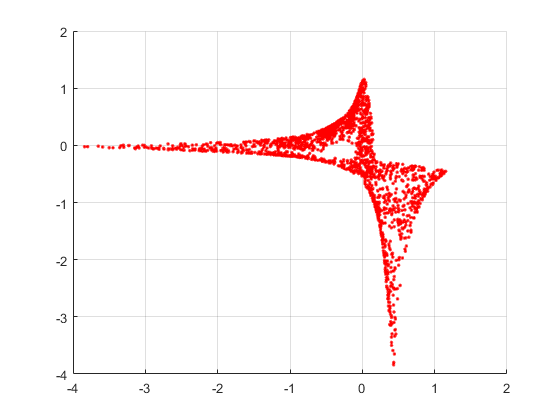}\\
\includegraphics[width=0.40\columnwidth]{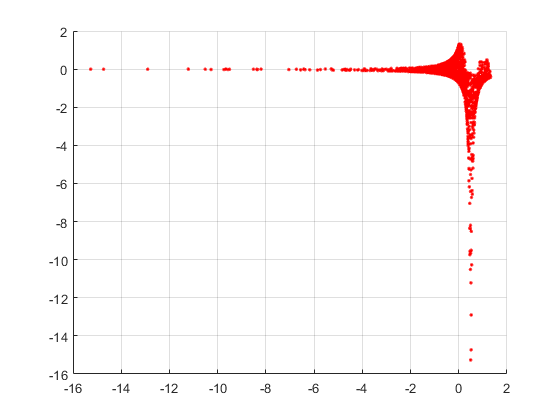}
\includegraphics[width=0.40\columnwidth]{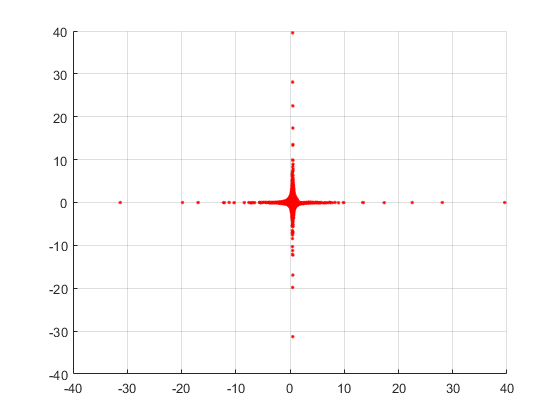}\\
\caption{First return maps of SBAF for $\alpha = 0.5$ and various values of $k < 0$ (top to bottom, left to right): $\{-2.0, -1.97, -1.90, -1.75, -1.5, -1.0 \}$ indicating absence of a fixed point. The plot confirms absence of fixed points for all values of $k<0$.}
\label{fig:firstreturnmaps1}
\end{center}
\end{figure}

\begin{figure}[htbp!]
\begin{center}
\includegraphics[width=0.40\columnwidth]{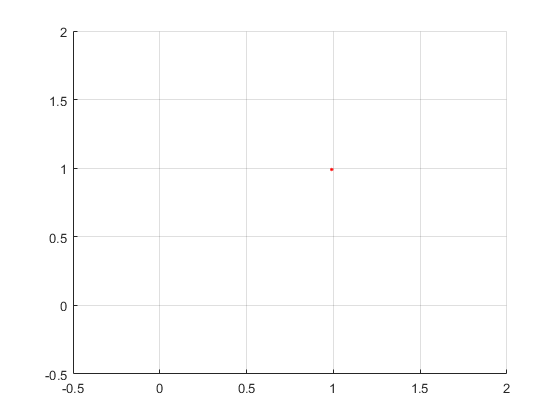}
\includegraphics[width=0.40\columnwidth]{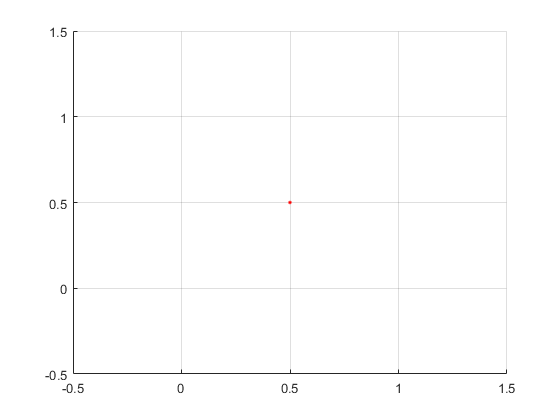}
\caption{First return maps of SBAF for $\alpha = 0.5$ and various values of $k > 0$: $\{0.1, 2.0 \}$ indicating presence of a fixed point.}
\label{fig:firstreturnmaps2}
\end{center}
\end{figure}

We can thus explicitly compute the values of $k$ for which there exists a stable fixed point when $\alpha = 0.5$:
\begin{align*}
kx^{*(0.5+1)}(1-x^*)^{-0.5}=1,
\end{align*}
which implies:
\begin{align*}
k =\frac{ \sqrt{(1-x^*)}}{x^*\sqrt{x^*}},
\end{align*}
where we seek the fixed point $x^*$ such that $0 <  x^* < 1$, $x^* = SBAF(x^*)$, and we also require for stability: $|SBAF'(x^*)|<1$. There are infinite number of stable fixed points satisfying these conditions. Numerically, we found that $x^*$ can be any value in the range $0<x^*<1$, and correspondingly, $k$ takes values: $\infty > k > 0$ (lower the $x^*$, higher the $k$). For example, if the stable fixed point is $x^*=\frac{1}{\sqrt{2}}$ then  $k = 0.91018$. When the fixed point varies from 0 to 1, the value of $SBAF'(x^*)$ varies from $-0.5$ to $0.5$ (thus $|SBAF'(x^*)|<1$, making it a stable fixed point).

Note that   We find it significant to mention that SBAF is inspired from a production function in economics \citep{Saha2016} even though the adaptation is non-trivial and significantly more complex in structure than the Cobb-Douglas production function, CDPF. Since CDPF is a production function defined by $k,\alpha, \beta$ and labor and capital inputs, a negative value of $k$ implies no quantity is produced, rather borrowing from market is necessary. This is important since it validates our choice of $k$ in neural net training from an econometric argument. This is consistent with our assertion that the choices of parameters for optimal classification performance are not accidental!

As noted above, stable fixed points exists for a range of values of $k$ depending on the value of $x^*$ and $\alpha$. When $\alpha = 0.5$, we found that there is a stable fixed point for every $0<k<\infty$. However, classification with SBAF works optimally at $k=0.91$ (corresponding $x^* \approx \frac{1}{\sqrt{2}}$). This is again consistent with our hypothesis that a stable fixed point will facilitate classification while chaos might occur otherwise. We have proven from the first order ODE that a fixed point exists and computed the fixed point in terms of $\alpha$ by fixing $k=1$ and verified the same via simulation. We have also observed (but not reported here) that altering $k$ values minimally within the stable fixed point range doesn't alter classification performance greatly. This reaffirms the confidence in the range of $k$ values obtained from fixed point analysis. To sum up, SBAF is the analytical solution to the first order DE and has a fixed point! This fixed point analysis is computationally verified and applied in the classification task on the PHl-EC data, via a judicious choice of parameters.

\subsection{Proof of global maxima for the activation}
We now go back to an economics perspective of our activation function (or production function). In this section, we prove the existence of an optimum solution.

Let us consider the activation function:
\begin{equation}
y = \frac{1}{1+kx^\alpha(1-x)^{1-\alpha}}
\end{equation}
$\alpha$ + $\beta$ =1 where $\alpha$ $>$ 0 and  $\beta >0$\\
Let  S = x and I = (1-x) and
1-$\alpha$ = $\beta$\\
For the constrained case of the above equation a critical point is defined in terms of the Lagrangian multiplier method. Suppose the constraint is
$g = w_1S + w_2I -m$\\
Let the Lagrangian function for the optimization problem be:
\begin{equation*}
    L = \frac{1}{1+kx^\alpha(1-x)^{1-\alpha}} - \lambda(w_1S + w_2I -m)\\
\end{equation*}
\begin{equation}
        L = \frac{1}{1+kS^\alpha I^{\beta}} - \lambda(w_1S + w_2I -m)
\end{equation}
On partially differentiating equation  (2) with respect to S,I  and $\lambda$ we obtain the following first order constraints
\begin{align*}
    \frac{\partial{L}}{\partial{S}} &= -yk\alpha S^{\alpha-1}I^{\beta} - \lambda w_1 =0 
    \\\frac{\partial{L}}{\partial{I}} &= -yk\beta S^{\alpha}I^{\beta-1} - \lambda w_2 =0
    \\\frac{\partial{L}}{\partial{\lambda}} &= -(w_1S + w_2I -m) = 0
\end{align*}
On differentiate again we obtain the second order condition:
\begin{align*}
    \\\frac{\partial{^{2}L}}{\partial{\lambda^{2}}} &= 0
    \\\frac{\partial{^{2}L}}{\partial{S^{2}}} &= -yk\alpha^{2} S^{\alpha-2}I^{\beta}  
    \\\frac{\partial{^{2}L}}{\partial{I^{2}}} &= -yk\beta^2 S^{\alpha}I^{\beta-2}
    \\\frac{\partial{^{2}L}}{\partial{SI}} &= -yk\alpha \beta S^{\alpha-1}I^{\beta-1}
    \\\frac{\partial{L}}{\partial{\lambda S}} &= -w_1 
    \\\frac{\partial{L}}{\partial{\lambda I}} &= -w_2 
\end{align*}
For equation (1) to obtain a optimum max value it subject to the assumed constraint ,it must satisfy the condition $\|M\| >$  0 where M is the bordered hessian matrix
\[
M=
  \begin{bmatrix}
    0 & -w_1 & -w_2 \\
    -w_1 & -yk \alpha^{2}S^{\alpha-2}I^{\beta} & -yk\alpha \beta S^{\alpha-1}I^{\beta-1} \\
   -w_2 & -yk\alpha \beta S^{\alpha-1}I^{\beta-1} & -yk\beta^2 S^{\alpha}I^{\beta-2}
  \end{bmatrix}
\]
$\|M\|$ = $w_1^{2}yk\beta^2 S^{\alpha}I^{\beta-2}$ +$w_1w_2yk\alpha \beta S^{\alpha-1}I^{\beta-1}$ + $w_2^{2}yk\alpha^2 S^{\alpha-2}I^{\beta}$ -$w_1w_2yk\alpha \beta S^{\alpha-1}I^{\beta-1}$ \\
$\|M\|$ = $w_1^{2}yk\beta^2 S^{\alpha}I^{\beta-2}$ + $w_2^{2}yk\alpha^2 S^{\alpha-2}I^{\beta}$\\
$\|M\| >  0 \rightarrow$ the production function has global optima under CRS.

\section{Approximating other activations} \label{sec:approximating}
In this section, we show the $k, \alpha$ values for which SBAF approximates other activation functions.

\begin{itemize}
    \item $k=1, \alpha= 0$; SBAF becomes $\frac{1}{2-x}$ which is what we obtain in sigmoid when we restrict the Taylor series expansion at 0 of exp$(-x)$ to first order!
    \item $k=1, \alpha= 1$; SBAF becomes $\frac{1}{1+x}$ which upon binomial expansion (restricting to first order expansion assuming $0<x<1$) yields $y= 1-x= 1-\text{ReLU}$
\end{itemize}

As noted earlier, these approximations can be seen in Figure \ref{fig:sbaf:2} along the top and bottom horizontal edges, which correspond to $\alpha=1$ and $\alpha=0$ respectively.

\subsection{ReLU approximate in the positive half} \label{sec:relu:approx:positivehalf}
Consider the function $f(x) = kx^n$.
We know that the ReLU activation function is $y = \max(0,x)$.
We need to approximate the values $n$ and $k$ such that $f(x)$ approximates to the ReLU activation function over a fixed interval. 

\begin{align*}
\text{Relative error} = \frac{||f(x) - y||}{||y||}
\end{align*}

Let the minimum tolerable error be $\epsilon < 10^{-3}$

\begin{align*}
    \frac{||f(x) - y||}{||y||} &<=\epsilon<10^{-3}\\
    ||f(x) - y||&\leq 10^{-3}||y||\\
    ||f(x)||&\leq ||y||(10^{-3} +1)\\
    ||f(x)||&\leq 1.001||y||\\
    ||f(x)||&\leq 1.001||y||\\
\end{align*}

Since $f(x) = kx^n$ approximates the positive half (i.e., when $x>0$) of the ReLU activation function, $y = max(x,0)$, the value of y when $x>0$ can be written as:

\begin{align*}
    ||y|| = ||x||
\end{align*}

Using this value in the error calculation we rewrite the error approximation as, 

\begin{align*}
    ||kx^n||&\leq 1.001||x||\\
    ||kx^{n-1}||&\leq 1.001
\end{align*}



The above is an optimization problem i.e. $\min \left\Vert kx^{n-1}\right\Vert$ subject to the constraints $k>0, n>1, -10<x<10$. We obtain the following bounds on $k$ and $n$:
\begin{align*}
    0<k<1; 
    1<n<2
\end{align*}
Therefore, we obtain the following continuous approximation of ReLU:
$y=kx^n, x>0$ when $0<k<1, 1<n<2, -10<x<10$, otherwise $y=0$. More precisely, the approximation to the order of $10^{-3}$ is $k=0.54, n=1.3$.
\begin{figure}
  \centering
  \includegraphics[scale=0.4]{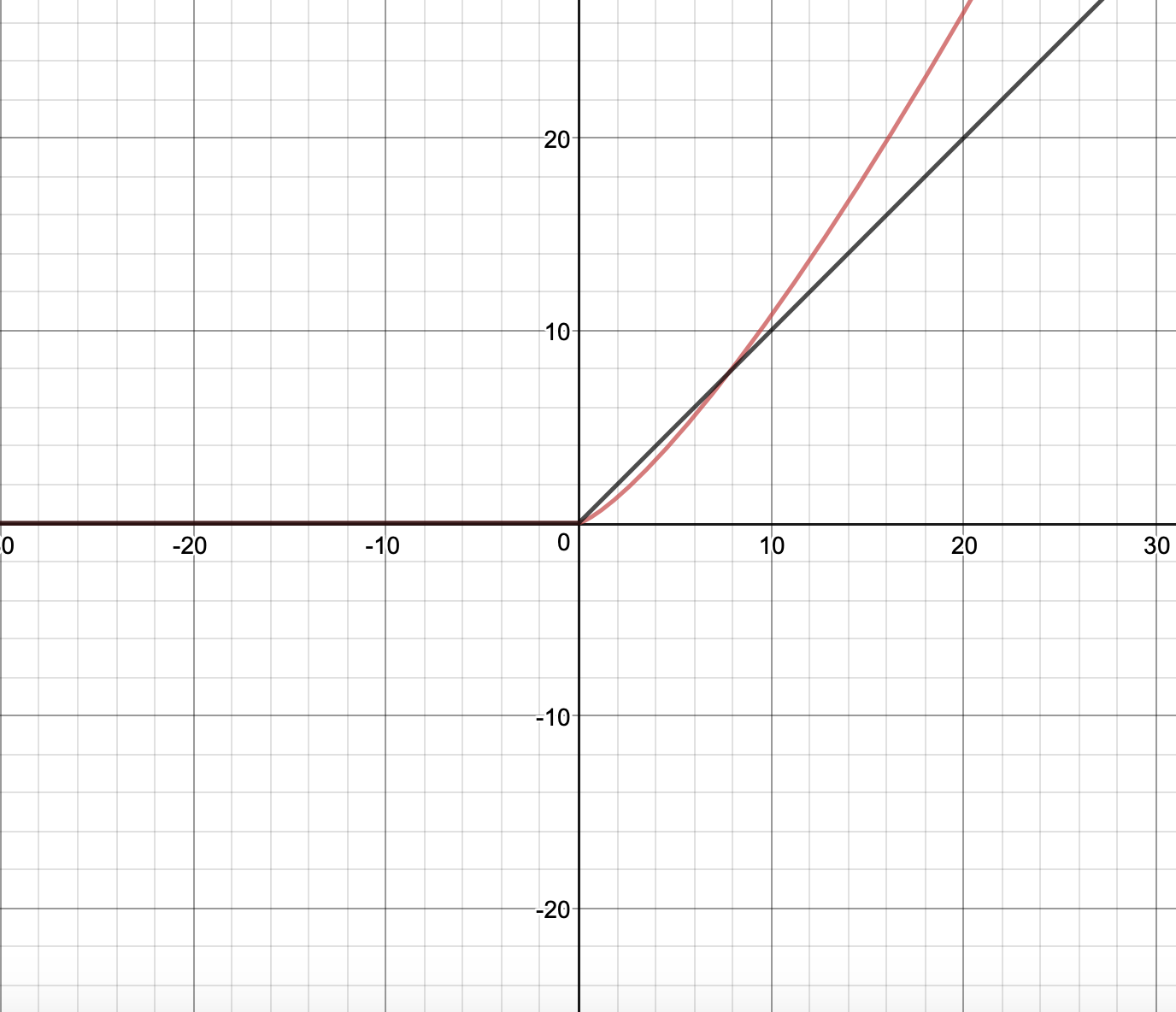}
  \caption{ReLU approximation $y=kx^n$. In $[0, 10]$, the values $k=0.54, n=1.30$, correct to two decimal places, yield the least approximation error as detailed in the text. The values of $k, n$ are used to train the network with A-ReLU. The two curves intersect at $(7.8, 7.8)$.}
  \label{fig:boat1}
\end{figure}

\subsection{Error approximation and estimation of k and n}
Define
\begin{equation}
L_2 = \sum||y_i-kx_i^n||^2
\end{equation}
We need to minimize the least square error to approximate the function $f(x) = ||y-kx^n||$ to the Relu activation function. This is a continuation in our efforts to find a continuous and differentiable approximation to ReLU.

Let us fix $x$ between some fixed interval, say $1<x<10$. This choice is also justified by observing linear combinations of weights and inputs during the training where we observed the combination rarely surpassing +3 in the positive half plane. This means $ x > 10$ would be least likely. Starting with some fixed value of $k$, say $k=0.5$, we try to approximate the value of $n$ such that the error in approximation is minimum. Algorithm \ref{alg:minkn} shows a simple way to achieve this.

\SetKw{KwBy}{by}
\begin{algorithm}[H]
\SetAlgoLined
min\_error $\gets \infty$\;
min\_k $\gets -1$, min\_n $\gets -1$\;
\For{$k \gets 0.5$ \KwTo $1$ \KwBy $0.01$}{
    \For{$n \gets 1$ \KwTo $2$ \KwBy $0.01$}{
        error $\gets 0$\;
        \For{$x \gets 1$ \KwTo $10$}{
            error $\gets$ sum $+ (kx^n-x)^2$\;
        }
        \If{error $<$ min\_error}{
            min\_error $\gets$ error\;
            min\_k $\gets k$\;
            min\_n $\gets n$\;
        }
    }
}
\textbf{return } $k, n$
\caption{Find $k, n$ with minimum error}
\label{alg:minkn}
\end{algorithm}

$f(x)$ is minimum at $k = 0.53$ or $k = 0.54$ when $n =1.3$. These are the parameter values used to train the network yielding better performance in classification, compared to ReLU. \\

\subsection{Differentiability of f(x)}
 \[   
 f(x) =
 \left\{
\begin{array}{ll}
      0 & x\leq0 \\
      kx^n & x > 0 \\
      
\end{array} 
\right. \]
We test for differentiability at $x = 0$. When $x \to 0^-$, $f^\prime(x) = 0$. When $x \to 0^+$, $f^\prime(x) = knx^{n-1} \to 0$. These derivatives are equal, and thus, the function is differentiable at $x=0$. Moreover, the derivative is continuous.

\subsection{Note about A-ReLU}
On plotting the curve for  the function, we obtain
\[   
 f(x) =
 \left\{
\begin{array}{ll}
      0 & x\leq0 \\
      kx^n & x > 0 \\
      
\end{array} 
\right. \] 
for different values of n and k, we have observed that the curve of f(x) for $x>0$ increases exponentially with increase in n. Since, the nature of the curve $kx^n$ is non linear when $k\neq0$, $k\neq1$ and $n\neq0$, $n\neq1$ it is not meaningful to compute the absolute error for the entire range on the positive scale from 0 to $\infty$. Hence, we have fixed the range of $x$ between 0 to $\infty$ on the positive scale and $-\infty$ to 0 on the negative scale. In section \ref{sec:relu:approx:positivehalf} we have found the relative error is bounded by $||x||$. In the chosen interval of $-\infty<x<10$, we find that the minimum absolute error  is $\approx$  $0.75\pm0.05$ when $k = 0.53$ or $k = 0.54$, and $n =1.3$. This value is quite small compared to our chosen bound, 10. Hence, we can say that the function $f(x)$ is a good approximation to the ReLU activation function. \footnote{Please note, even if we don't restrict the function in the specified range mentioned above, we can work with the function itself as another activation function with no discontinuity at $x=0$. We choose the value of $n$ between 1 and 2 so that the derivative doesn't explode!}

For a continuous domain, the squared error changes from a sum to an integral. We take this integral over $(0, t)$, for some $t$ that we choose later. To minimize this, we take the partial derivatives with respect to $k$ and $n$, and set them to 0. This yields two equations, with two unknowns. $t$ can be thought of as the upper limit of the domain where the approximation is good.
    
    \begin{align*}
        f(k, n) &= \int_0^{t} (kx^n-x)^2 \mathrm{d}x \\
                &= \int_0^{t} \left( k^2 x^{2n}+x^2 -2kx^{n+1} \right) \mathrm{d}x \\
                &= k^2 \frac{t^{2n+1}}{2n+1}+\frac{t^3}{3}-\frac{2k t^{n+2}}{n+2} \\
        \frac{\partial f}{\partial k} &= 2k \frac{t^{2n+1}}{2n+1}-\frac{2t^{n+2}}{n+2} = 0 \\
                &\Rightarrow k\frac{t^{2n+1}}{2n+1} = \frac{t^{n+2}}{n+2} \\
                &\Rightarrow \boxed{k = \left( \frac{2n+1}{n+2}\right) t^{1-n}} \\
        \frac{\partial f}{\partial n} &= k^2 \left( \frac{2 t^{2n+1}\ln (2n+1)}{2n+1} - \frac{2t^{2n+1}}{(2n+1)^2} \right) - 2k \left( \frac{t^{n+2}\ln(n+2)}{n+2}-\frac{t^{n+2}}{(n+2)^2} \right) = 0 \\
                &\Rightarrow \boxed{k \left( \frac{t^{2n+1}\ln (2n+1)}{2n+1} - \frac{t^{2n+1}}{(2n+1)^2} \right) = \frac{t^{n+2}\ln(n+2)}{n+2}-\frac{t^{n+2}}{(n+2)^2}}
    \end{align*}

    \begin{figure}
        \centering
        \includegraphics[scale=0.4]{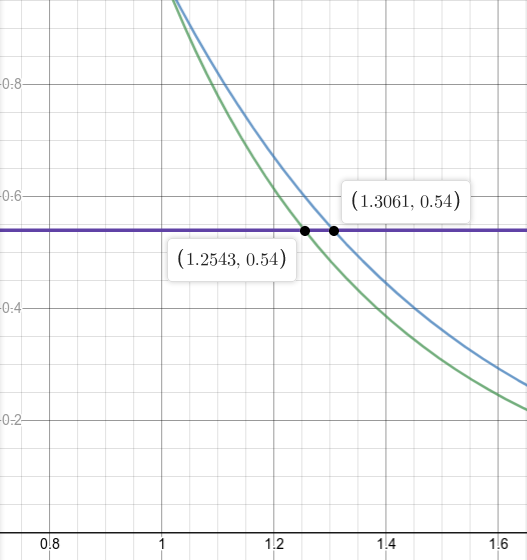}
        \caption{Plot of the two curves derived analytically, showing the intersection with $k=0.54$. The blue curve is the first one, and the green curve is the second.}
        \label{fig:analytical:1}
    \end{figure}

    Because it is difficult to a solution to this system of equations analytically, we simply plot these equations, fixing $t=10$. In the graph below, the y-axis is $k$, and the x-axis is $n$. Note that the intersection of the two equations yields the trivial solution $k=n=1$; however, this solution is not very interesting. Instead, we fix the value of $k$, and find the intersections with the two curves above. Figure \ref{fig:analytical:1} shows this plot.

\section{Network architecture}
The architecture to explore the proposed activation functions is based on a multilayer perceptron that internally deploys back-propagation algorithm to train the network. Neurons are fully connected across all the layers and output of one layer is supplied as input to neurons of subsequent layer during the forward pass. Alternatively, the backward pass computes the error gradient and transmits it back into network in the form of weight-adjustments. Computation of gradients require calculating the derivatives of activation functions (SBAF and A-ReLU) which have been explained briefly in Algorithm 2 and 3. 

The implementation of these algorithms is done in Python installed on an x64 based AMD E1-6010 processor with 4GB RAM. The github repository\footnote{\url{https://github.com/mathurarchana77/A-RELUandSBAF}} stores the Python code for  the SBAF and A-ReLU activations. The purpose of this exercise is to demonstrate the performance of the activation functions used on variety of data sets. The Python implementation of these activation functions is done from scratch. The whole architecture is implemented as a nested list, where the network is stored as a single outer list and multiple layers in the network are maintained as inner lists. A dictionary is used to store connection weights of the neurons, their outputs, the error gradients and other intermittent results obtained during back propagation of errors. The computations associated with the processing of neurons in the forward and backward pass are indicated in the algorithms below. The next sections explores the details involved in  execution of these algorithms on different feature sets and investigates the performances by comparing them with the state-of-the-art activation functions.

\SetKw{KwBy}{by}
\begin{algorithm}[H]
\SetAlgoLined
Initialize all weights $w_{ij}$, biases $b_i$, $n\_epochs$, lr, k and $\alpha$\;
\For{Each training tuple $I$ in the database}{
    \For{each unit $j$ in input layer}{
    $O_j \gets I_j$\;
    }
    The forward Pass\;
    \For{ each unit $j$ in the hidden layer}{ 
    $h_{jnet} \gets \sum_{i}w_{ji} O_i + b_j$\;
    $h_{jout} \gets \frac{1}{1+k (h_{jnet})^\alpha (1-h_{jnet})^{(1- \alpha)}} $\;
    }
\For{ each unit $j$ in the output layer}{
$o_{jnet} \gets \sum_{i}w_{ji} h_{iout} + b_j $\;
$o_{jout} \gets \frac{1}{1+k (o_{jnet})^\alpha (1-o_{jnet})^{(1- \alpha)}} $\;
}

The Backward pass\;
\For{each unit $j$ in the output layer}{ 
$d \gets \frac{o_{jout}(1-o_{jout})}{o_{jnet}(1-o_{jnet})} (o_{jnet}-\alpha)$\;
 $E \gets d . (T_j - O_j)$\; 
}
\For{each unit $j$ in the hidden layer}{
$d \gets \frac{h_{jout}(1-h_{jout})}{h_{jnet}(1-h_{jnet})} (h_{jnet}-\alpha)$\;
$E \gets d  \sum_{k} E_k W_{kj}$\; 
}
The weights update\;
\For{each weight $w_{ji}$ in network}{ 
$ \triangle w_{ji} \gets lr . E_j. h_{jout}$\;
$w_{ji} \gets w_{ji} + \triangle w_{ij}$\;
}
The bias update\;
\For{each weight $b_{i}$ in network}{ 
$ \triangle b_{i} \gets lr . E_j$\;
$b_{i} \gets b_{i} + \triangle b_{i}$\;
}
}
\caption{Optimise weights and biases by using back propagation on SBAF}

\end{algorithm}

\newpage
\SetKw{KwBy}{by}
\begin{algorithm}[H]
\SetAlgoLined
Initialize all weights $w_{ij}$, biases $b_i$, $n\_epochs$, lr, k and n\;

\For{Each training tuple $I$ in the database}{
\For{each unit $j$ in input layer}{
$O_j \gets I_j$ \;
}

\For{ each unit $j$ in the hidden layer}{
$h_{jnet} \gets \sum_{i}w_{ji} O_i + b_j $ \;
$h_{jout} \gets k.(h_{jnet})^n$ \;
}
\For{ each unit $j$ in the output layer}{
$o_{jnet} \gets \sum_{i}w_{ji} h_{iout} + b_j $\;
$o_{jout} \gets k.(o_{jnet})^n $;\
}

\For{each unit $j$ in the output layer}{
$d \gets n.k^\frac{1}{n}.(o_{jout})^\frac{n-1}{n}$\;
$E \gets d . (T_j - O_j)$\;
}
\For{each unit $j$ in the hidden layer}{
$d \gets n.k^\frac{1}{n}.(h_{jout})^\frac{n-1}{n}$\;
$E \gets d  \sum_{k} E_k w_{kj}$\;
}
\For{each weight $w_{ji}$ in network}{
$ \triangle w_{ji} \gets lr . E_j. h_{jout}$ \;
$  w_{ji} \gets w_{ji} + \triangle w_{ij}$ \;
}
\For{each weight $b_{i}$ in network}{
$ \triangle b_{i} \gets lr . E_j$ \;
$  b_{i} \gets b_{i} + \triangle b_{i}$\;
}
}
\caption{Optimise weights and biases by using back propagation on A-ReLU}
\end{algorithm}



\section{Data}
The PHL-EC (University of Puerto Rico's Planetary Habitability Laboratory's Exoplanet Catalog) dataset \citep{phlref,SchulzeMakuch2011} consists of a total of 68 features, 13 categorical and the remaining 55 are continuous. The catalog uses stellar data from the Hipparcos catalog \citep{hipparcosref} and lists $3771$ confirmed exoplanets (at the time of writing this paper), out of which $47$ are meso and psychroplanets and the remaining are non-habitable. The catalog includes important features like atmospheric type, mass, radius, surface temperature, escape velocity, Earth Similarity Index, flux, orbital velocity etc. The difference between The PHL-EC and other catalogs is that PHL-EC models some attributes when data are not available. This includes estimating surface temperature from equilibrium temperature as well as estimating stellar luminosity and pressure. The presence of observed and estimated attributes presents interesting challenges.

This paper uses the PHL-EC data set, of which an overwhelming majority are non-habitable samples. Primarily, PHL-EC class labels of exoplanets are non-habitable, mesoplanets, and psychroplanets \cite{phlref}. The class imbalance is observed in the ratio of thousands to one. Further, the potentially habitable planets (meso or psychroplanets) have their planetary attributes in a narrow band of values such that the margins between mesoplanets and psychroplanets are incredibly difficult to discern. This poses another challenge to the classification task.

The classes in the data are briefly described below:
\begin{enumerate}
\item \textbf{Non-Habitable}: Planets, mostly too hot or too cold, may be gaseous, with non-rocky surfaces. Such conditions don't favor habitability.

\item \textbf{Mesoplanet}: Generally referred to as Earth-like planets, they have sizes between that of Ceres (the largest minor planet in our solar system) and Mercury. The average global surface temperature is usually between 0$^\circ$C and 50$^\circ$C. However, Earth-similarity is no guarantee of habitability.

\item \textbf{Psychroplanet}: These planets have mean global surface temperature between -50$^\circ$C to 0$^\circ$C. Temperatures of psychroplanets are colder than optimal for sustenance of terrestrial life, but some psychroplanets are still considered as potentially habitable candidates.
\end{enumerate}

The data set also has other classes, though insignificant in number of samples for each class. These are thermoplanets, hypopsychroplanets and hyperthermoplanets. The tiny number of samples in each class makes it unsuitable for the classification task and these classes are therefore not considered for class prediction. Certain features such as the name of the host star and the year of discovery have been removed from the feature set as well. We filled the missing data by class-wise mean of the corresponding attribute. This is possible since the amount of missing data is about $1\%$ of all the data. The online data source for the current work is available at the university website  \footnote{http://phl.upr.edu/projects/habitable-exoplanets-catalog/data/database}.

\section{Experiments}
The PHL-EC dataset has 3771 samples of planetary data for 3 classes of planets and 45 features (after pruning unnecessary features). As already mentioned, a Multi Layered Perceptron (MLP) is implemented at the core for classifying planets. The MLP internally utilizes  gradient descent  to update  weights and biases during classification. The connection weights and biases are randomly initialized. The activation functions used in MLP are sigmoid, SBAF, A-ReLU and ReLU, and details for implementing SBAF and A-ReLU along with the computation of gradients in both cases is shared in the Appendix. 

As a part of the experiments, different data sets are explicitly  built  by selecting certain combination of features from the original feature set. The idea behind doing this is to evaluate the performance of functions for a variety of feature sets. The following subsection illustrates various sets  of data used on the activation functions and their results. In all these cases, the training set consists of 80\% of samples and remaining 20\% are used for testing. SBAF uses the hyper parameters,  $\alpha$ and $k$, that are tuned during execution of code and  best results were seen at $\alpha$ = 0.5 and k = 0.91 (in agreement with the fixed point plots we observed, $k=1$ doesn't alter classification performance). Similarly, for A-ReLU, $k$ and $n$ were set to 0.5 and 1.3 respectively (this is from the evidence by the approximation to ReLU-empirical observations match). This eliminates the need for parameter tuning. 

The results of classification (Tables 2 and 3) are interesting. The accuracy, precision and recall is indicated for all classes of planets: Non habitable, Mesoplanet and Psychroplanet. As seen from the tables, A-ReLU has outperformed ReLU in almost all the cases under investigation and SBAF has performed significantly better than the parent function, Sigmoid. 

\subsection{Chosen feature sets} 
The different combination of features  employed on the traditional and  proposed activation functions is explored here.  A case-by-case exploration of these feature sets with their performance comparison, is indicated in Tables 2 and 3. The first column of these tables reveal features, marked with their count and the case number. Remaining columns reflect class-wise accuracy, precision and recall of the 4 activation functions.

\begin{enumerate}
    \item To begin with, all features are used as input to the neural network, thus leading to 45 input and 3 output neurons. Since the inherent characteristics of each activation function is different, each one uses a different number of neurons in the hidden layer to reach convergence. For sigmoid, 20 hidden neurons are used while SBAF, A-ReLU and ReLU used 11 neurons in the hidden layer. Sigmoid gives the best accuracy at learning rate of 0.015, momentum of 0.001 and at 500 epochs, while SBAF, A-ReLU and ReLU gives the best results at 100 epochs keeping the other parameters same.
    \item A subset of features (restricted features) consisting of Planet Minimum Mass, Mass,	Radius,	 SFlux Minimum, SFlux Mean, SFlux Maximum are used as input.  All networks used 4 neurons in hidden layer to stabilize. Sigmoid converged at learning rate of 0.1  and momentum of 0.01. In parallel, SBAF  A-ReLU and ReLU converged at learning rate of 0.08, momentum at 0.004 and number of epochs as 300. The performance of all the classifiers  is reported in Table 2. It is evident from the table that the network is able to classify even the minority class samples (Mesoplanets and Psychroplanets) with fairly decent accuracy.
    \item It has already being shown that Surface Temperatures (ST) can  distinguish habitable planets from non-habitable ones with a large degree of precision. Therefore, it becomes essential to ascertain if the proposed activation functions (SBAF and A-ReLU) as well as the already established ones (sigmoid and ReLU), can perform classification when ST is removed from feature set. To achieve this, the next set of features are those from which  ST is removed. Thus, exoplanet features used as input are Zone Class, Mass Class,Composition Class, Atmosphere Class,Min Mass, Mass, Radius, Density, Gravity, Esc Vel ,SFlux Min, SFlux Mean, SFlux Max, Teq Min (K)	, Teq Mean (K),Teq Max (K), Surf Press, Mag, Appar Size, Period, Sem Major Axis, Eccentricity, Mean Distance , Inclination, Omega , HZD, HZC, HZA, HZI,  ESI and Habitable. The features of parent Star that belong to feature set are Mass, Radius, Teff, Luminosity,	Age, Appar Mag,	Mag from Planet,	 Size from Planet,	 Hab Zone Min and Hab Zone Max. For SBAF and A-ReLU, the 44 featured data set is tuned at learning rate = 0.01, momentum = 0.001 and epochs = 300. The number of units in the hidden layer is 12 for all four activation functions. It is interesting to inspect that A-ReLU performs 100\% classification for all three classes and this is at par with A-ReLU. However, sigmoid performs marginally better than SBAF.
    \item A unique combination of planetary attributes like  Minimum Mass, Mass, Radius and Composition class are taken as input. The essence of selecting these features is to know whether the activation functions can perform classification by solely using Mass and mass related features. It becomes a challenging machine learning problem since the discrimination of planets is not at all clear by features such as mass of planets. Interestingly, for this kind of problem, A-ReLU performs better than the parent function ReLU, and SBAF outpaces Sigmoid with substantial difference in accuracy. The networks are tuned at learning rate of 0.2 and momentum of 0.03 at 400 epochs. It uses 3  neurons in the hidden layer.
    \item The following 4 cases (Cases 5-8) are set for an exclusive exploration, where exoplanets are classified using one  feature from planets at a time alo ng with multiple features of the parent star. At first, the planet's radius along with 6 star features (Mass, Radius, Teff, Luminosity, Hab Zone Min and Hab Zone Max) are fed as input to the four activation functions. With 4  neurons in the hidden layer, the learning rate and momentum for the network is tuned to 0.09 and 0.001. Results are achieved at 300 epochs, as shown in Table 3. The performance of A-ReLU is exceedingly better than the rest of the activation functions.
    \item Continuing on the same line of work, another feature set bearing planet's mass and previously used 6 star features are fed to the network. Hidden neurons are same in number as in the previous case. Here too, A-ReLU has performed better from its parent activation function, ReLU as well as from the other functions in terms of classification accuracy, precision and recall.
    \item Table 3 shows the performance of classification using Minimum mass as planet's feature and remaining parent star features as input keeping other arrangements same as  the previous cases. Comparing  accuracy in each activation function, A-ReLU has again performed better for all the 3 classes of planets.
    \item A slightly different grouping of feature is attempted here. Two planet features, mass and minimum distance computed from parent star, and the remaining 6 star features are used as input. For this particular combination of features, A-ReLU and ReLU performs at par and sigmoid performs better classification in comparison with the rest. This is the only case where an irregularity is seen in the performance.
\end{enumerate}

The performance metrics shown in Table 2 and 3 indicates that SBAF and A-ReLU outshine sigmoid and ReLU in almost all of the cases. Some more critical parameters in terms of training time, CPU utilization and memory requirements, for the execution of activation functions is reported in Table 4. SBAF and A-ReLU take the shortest time to reach convergence even though number of epochs are kept same. CPU utilization is minimum for A-ReLU followed by ReLU, SBAF and sigmoid. Memory consumption is balanced and almost equal for all the functions under study. It is worth mentioning that values reported in the table are obtained after taking the average of all 8 executions from above cases.

An investigation of the confusion matrix resulting from the execution of SBAF, is disclosed in Table 1. It is noted that, even though there are considerably large number of samples of non-habitable planets, SBAF is able to classify non-habitable and Psychroplanet unmistakably, with the accuracy of 1 and 0.994 for the respective classes. However, Mesoplanets are not flawlessly classified, (out of 9 samples, 5 were correctly labeled and rest of the 4 were mistakenly labeled to be of Psychroplanet), the reason can be attributed to the class distribution in the data set. The number of samples of non-habitable, Mesoplanet and Psychroplanet  are 3724, 17 and 30 respectively. Evidently, Mesoplanet are lowest in number  and thus, the number of samples were not sufficient to train the network. Nevertheless, this can be handled  by generating synthetic data for balancing the class samples. 

\begin{table}[h]
\centering
\begin{tabular}{l|l|c|c|c|c}
\multicolumn{2}{c}{}&\multicolumn{3}{c}{Predicted}&\\
\cline{3-5}
\multicolumn{2}{c|}{}&Non-Habitable&Mesoplanet&Psychroplanet\\
\cline{2-5}
\multirow{2}{*}{Actual}& Non-habitable & 745 & 0 & 0\\
\cline{2-5}
& Mesoplanet & 0 & 5 & 4\\
\cline{2-5}
& Psychroplanet & 0 & 0 & 1\\
\cline{2-5}
\end{tabular}
\caption{Confusion Matrix obtained while executing SBAF using all 45 features;the three classes are non habitable, Mesoplanet and Psychroplanet; the confusion matrix shows all 745 non-habitable planets classified correctly; 5 Mesoplanets and 1 Psychroplanet is also labeled correctly by the network. Please note, all rocky exoplanets have been considered and therefore the numbers don't match up with the total count reported in introduction.}
\end{table}

\begin{table*}[!htbp]
\centering
\resizebox{\textwidth}{!}{
\begin{tabular}{| c | c | c c c | c c c | c c c | c c c | c c c |}
\hline
\multirow{4}{*}{Different feature sets} & & \multicolumn{15}{c|}{Different Activation functions used in the study} \\ \cline{3-17}
 
 & \multirow{2}{*}{Performance} & \multicolumn{3}{c|}{Sigmoid} & \multicolumn{3}{c|}{SBAF} & \multicolumn{3}{c|}{Approx. Relu} & \multicolumn{3}{c|}{Relu} & \multicolumn{3}{c|}{Leaky ReLU} \\ \cline{3-17}

 &  & Non-H & Meso & Psychro & Non-H & Meso & Psychro & Non-H & Meso & Psychro & Non-H & Meso & Psychro & Non-H & Meso & Psychro\\
\hline

& Accuracy & 1.    &     0.975 & 0.975  & 1.0& 0.99& 0.994 & 0.997 & 1.   &      0.997  &  1.0 & 0.99 & 0.99 & 0.994 & 0.5 & 0.857 \\
All features (45) & Precision &  1.0  & 0.943 & 0.932 & 1.0& 1.0& 0.2 & 1.    &     1.  &       0.987 & 0.988 &  0.998 &  & 1.0 & 0.25 & 0.857 \\
(Case 1) & Recall &  1.0     & 0.895 & 0.965  & 1.0& 0.55& 1.0  &  0.997 & 1.0   &      1.0    & 0.998  & 0.991 &  0.991 & 0.994 & 0.5 & 0.857 \\
\hline
Restricted & Accuracy &  0.758 & 0.741 & 0.798  & 0.987& 0.854&  0.847 &  1.0 & 1.0  & 1.0   &  0.985 & 0.834 & 0.830 & 0.937 & 1.0 & 0.286 \\
 Features (6) & Precision & 0.719 & 0.525 & 0.808 & 1.  &       0.681 & 0.643 & 1.0  &       1.0  & 1.0   &  1.0 & 0.5 & 0.618 & 1.0 & 0.18 & 0.2 \\
(Case 2) & Recall & 0.864 &  0.583&  0.384 & 0.978 & 0.223  & 0.943 & 1.0  &1.0   & 1.0   &  0.974 & 0.149 & 0.925 & 0.937 & 1.0 & 0.286 \\
\hline
Without  Surface& Accuracy & 0.998 & 0.998 & 0.997 & 0.997 & 0.924 & 0.927 & 1.0 & 1.0  & 1.0   &  1.0 & 1.0 & 1.0 & 0.98 & 1.0 & 0.67 \\
Temperature (44)  & Precision & 0.994 & 0.995 & 1.         & 1.  &       0.484 & 0.783  &  1.0 & 1.0  & 1.0   & 1.0 & 1.0 & 1.0 & 1.0 & 0.5 & 0.33 \\
 (Case 3)& Recall & 1.   &      1.  &       0.994 & 0.996 &  0.5   &     0.783 &  1.0 & 1.0  & 1.0   & 1.0 & 1.0 & 1.0 & 0.98 & 1.0 & 0.67 \\
\hline
Planet MinMass & Accuracy &  0.888 & 0.825 & 0.750 & 0.922 & 0.922 & 0.864 & 0.971 & 0.946 & 0.931 &  0.968 & 0.937 & 0.905 & 0.976 & 0.67 & 0.71 \\
Mass and  & Precision &  0.974 & 0.534 & 0.547 & 1.  &    0.3   &  0.456  & 1.    &     0.619 &0.673   &  0.996  &    1. &        0.567 & 0.997 & 0.25 & 0.56 \\
Radius (4) (Case 4) & Recall &  0.807 & 0.397 & 0.814  & 0.904 & 0.130 & 0.912&  0.965 & 0.590 & 0.846   &  0.965 & 0.090 & 0.974 & 0.976 & 0.67 & 0.71 \\
\hline
\end{tabular}}
\caption{Performance analysis of different Activation functions used in the study. First column indicates features that are given as input, along with their count. 3rd, 4th, 5th, 6th, and 7th columns show the performance of 5 different activation functions. Accuracy, precision and recall is indicated for every class of planet - Non habitable, Mesoplanet and Psychroplanet. A-ReLU has out performed ReLU, LeakyReLU and Sigmoid by significant difference in performance. Difference in performance is clearly visible in minority classes, Meso and psychro. Note, SBAF and A-ReLU didn't need parameter tuning, at all.}
\label{table:big:1}
\end{table*}

\begin{table*}[!htbp]
\centering
\resizebox{\textwidth}{!}{
\begin{tabular}{| c | c | c c c | c c c | c c c | c c c |c c c|}
\hline
\multirow{4}{*}{Different feature sets} & & \multicolumn{15}{c|}{Different Activation functions used in the study} \\ \cline{3-17}
 
 & \multirow{2}{*}{Performance} & \multicolumn{3}{c|}{Sigmoid} & \multicolumn{3}{c|}{SBAF} & \multicolumn{3}{c|}{Approx. Relu} & \multicolumn{3}{c|}{Relu} & \multicolumn{3}{c|}{Leaky ReLU} \\ \cline{3-17}

 &  & Non-H & Meso & Psychro & Non-H & Meso & Psychro & Non-H & Meso & Psychro & Non-H & Meso & Psychro & Non-H & Meso & Psychro \\
\hline
Radius & Accuracy & 0.848 & 0.753 & 0.608 & 0.721 & 0.738 & 0.754 & 0.908 & 0.8311 & 0.807 &  0.890 & 0.861 & 0.795 & 0.813 & 0.5 & 0.57 \\
and other & Precision & 0.856 & 0.668 & 0.440 & 0.617 & 0.669 & 0.152 & 0.919 & 0.718 & 0.548 & 0.883 & 0.749 & 0.516 & 1 & 0.07 & 0.03 \\
Star features (Case 5) & Recall &  0.655 & 0.515 & 0.645  & 0.962 & 0.738 & 0.37  &  0.867 & 0.852 & 0.441   & 0.866 &  0.908 & 0.354 & 0.813 & 0.5 & 0.57 \\
\hline
 Planet Mass & Accuracy & 0.846 & 0.706 & 0.58 & 0.859 & 0.868 & 0.881 & 0.997 & 1.   &      0.997  & 0.877& 0.899& 0.902 & 0.837 & 0.71 & 0.5 \\
and other& Precision & 0.764 & 0.55 & 0.337 & 0.914 & 0.442 & 0.307 & 1.    &     1.  &       0.987 &  0.878 & 0.554 & 0.0 & 0.997 & 0.116 & 0.045 \\
Star features (Case 6) & Recall & 0.78 & 0.65 & 0.27 & 0.906 & 0.628 & 0.177  &  0.997 & 1.   &      1.    &  0.980 & 0.580 & 0.0 & 0.837 & 0.71 & 0.5 \\
\hline
Minimum & Accuracy & 0.85 & 0.683 & 0.7 & 0.850 & 0.8007 & 0.857 & 0.925 & 0.850 & 0.868  & 0.864 & 0.893 & 0.864 & 0.88 & 1 & 0\\
Mass and other & Precision & 0.711 & 0.75 & 0.532 & 1.0 & 0.190 & 0.520 & 0.938 & 0.269 & 0.571 &  0.857&       0.0 & 0.558 & 1 & 0.045 & 0 \\
Star features (Case 7) & Recall & 0.925 & 0.07 & 0.85 & 0.798 & 0.2666 & 0.8837 & 0.961 & 0.233 & 0.558 &0.980 & 0.0& 0.558 & 0.88 & 1 & 0 \\
\hline
Minimum Mass & Accuracy &  0.958 & 0.8 & 0.808  & 0.858 & 0.77 & 0.716 & 0.904 & 0.799 & 0.839  &  0.901& 0.796 & 0.790 & 0.87 & 1 & 0 \\
P. Distance and & Precision &  1.&  0.785 & 0.649  & 0.911 & 0.35 & 0.38 &0.894 & 0.506 & 0.666 & 0.893 &0.000 & 0.478 & 1 & 0.05 & 0 \\
Star features (Case 8) & Recall &  0.875 & 0.55 & 0.925  & 0.846 & 0.106 & 0.746  &  0.958 & 0.636 & 0.349 & 0.9485 & 0.0    & 0.888 & 0.868 & 1 & 0   \\
\hline
\end{tabular}}
\caption{Performance analysis of different Activation functions used in the study. First column indicates input feature along with their count. 3rd, 4th, 5th and 6th column shows the performance of 4 different activation functions. Accuracy, precision and recall is indicated for every class of planet - Non habitable, Mesoplanet and Psychroplanet. A-ReLU has outshined significantly, the parent activation function ReLU in terms of performance. An anomaly is seen for case 8 where sigmoid performed marginally better than SBAF.}
\label{table:big:2}
\end{table*}

\begin{table*}
\begin{center}
\begin{tabular}{ |c|c|c|c|c|c| } 
\hline
\multirow{2}{*}{} &  \multicolumn{5}{c|}{ Activation functions (Epochs =500) } \\ \cline{2-6}
& Sigmoid & SBAF & A-ReLU & ReLU & Leaky ReLU \\
\hline
Learning rate, momentum & 0.01, 0.001 & 0.01, 0.01  & 0.01, 0.001 & 0.01, 0.01 & 0.01, 0.001 \\ 
\hline
Time(seconds) & 409.33 &281.97  & 312.31& 358.01 & 574.6 \\ 
\hline
CPU Utilization (\%) & 54.6 & 53.6 & 42.8 & 48.4 & 41.81 \\
\hline
Memory Usage (MB) & 67.7 & 67.8 & 67.4 & 67.8 & 45.45 \\
\hline
\end{tabular}
\end{center}
\caption{Analysis of Training time, CPU utilization and Memory usage for Sigmoid, SBAF, A-ReLU and ReLU at epochs = 500. The reported values are averaged over all 8 executions mentioned above. The  SBAF has the lowest training time followed by A-ReLU. CPU utilization is minimum for A-ReLU; SBAF has marginally better CPU utilization from the parent activation function sigmoid.   }
\end{table*}

\begin{figure}[htbp!]
\centering

\begin{subfigure}[b]{0.4\textwidth}
    \centering
    \includegraphics[width=\textwidth]{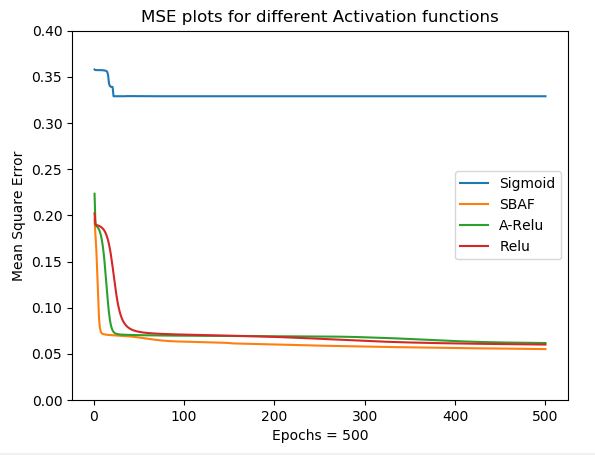}
\end{subfigure}
\begin{subfigure}[b]{0.4\textwidth}
    \centering
    \includegraphics[width=\textwidth]{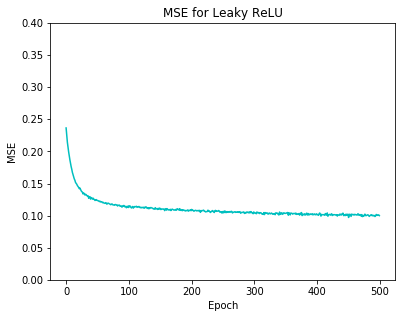}
\end{subfigure}
\caption{MSE plots for all Activation functions executed for 500 epochs on Restricted features of PHL-EC data: Sigmoid attains saturation at an early MSE value and fails to learn effectively when compared with SBAF, A-ReLU and ReLU. This fact may be interpreted as the representational power of the activation function (since identical network architecture was used). If the activation function is more expressive, it will, in general, get a lower MSE when trained sufficiently long enough. Therefore, it's not surprising that SBAF has the lowest MSE (and thus the highest representational power), since its shape can easily be changed. A-ReLU shouldn't be very surprising either since the parameters render A-ReLU flexibility. Even though SBAF and A-ReLU has more parameters compared to sigmoid, their faster convergence is striking. }
\label{fig:mse:3}
\end{figure}

\newpage
Even though our focus is not all on the performance of statistical machine learning or other learning algorithms on the PHL-EC data, we feel it's necessary to report the performance of some of those briefly. This was a post-facto analysis (the realization dawned upon us much later) done to ascertain the efficacy of our methods compared to methods such as Gaussian Naive Bayes (GNB), Linear Discriminant Analysis (LDA), SVM, RBF-SVM, KNN, DT, RF and GBDT \citep{Insight}. These methods didn't precede the exploration into activation functions. We believe the readers should know and appreciate the complexity of the data and classification task at hand, in particular when hard markers (surface temperature and surface temperature related features which discriminate the habitability classes quite well). This also highlights the remarkable contribution of the proposed activation functions toward  performance metrics. It is for this reason, we don't tabulate the results in detail but just state that for the specific cases (Cases 2, 4, 5-8, "six" out of the total 'eight" cases considered for our experiments) where "hard marker" features were removed, none of the methods mentioned above reached over 75\% accuracy class-wise and 68\% overall. This augurs well for the strength of our analysis presented in the manuscript.

\section{Discussion and Conclusion}

The motivation of SBAF is derived from the idea of using $kx^{\alpha}(1-x)^{1-\alpha}$ to maximize the width of the two separating hyperplanes (similar to separating hyperplanes in the SVM as the kernel has a global optima) when $0 \leq \alpha \leq 1$. This is equivalent to the CDHS formulation when CD-HPF is written as $ y = kx^{\alpha}(1-x)^{\beta}$ where $\alpha+\beta=1, 0 \leq \alpha \leq 1, 0 \leq \beta \leq 1$, $k$ is suitably assumed to be $1$ (CRS condition), and the representation ensures global maxima (maximum width of the separating hyperplanes) under such constraints \citep{CDHPF2016,Potential}. 
The new activation function to be used for training a neural network for habitability classification boasts of an optima. Evidently, from the graphical simulations presented earlier, we observe less flattening of the function and therefore the formulation is able to tackle local oscillations more easily as compared to the more generally used sigmoid function. Moreover, since $0 \leq \alpha \leq 1, 0 \leq x \leq 1, 0 \leq 1-x \leq 1$, the variable term in the denominator of SBAF, $kx^{\alpha}(1-x)^{1-\alpha}$ may be approximated to a first order polynomial. This may help us in circumventing expensive floating point operations without compromising the precision. We have seen evidence of these claims, theoretically and from implementation point of view, in the preceeding sections.

Habitability classification is a complex task. Even though the literature is replete with rich and sophisticated methods using both supervised \citep{Zighed2010} and unsupervised learning methods, the soft margin between classes, namely psychroplanet and mesoplanet makes the task of discrimination incredibly difficult. A sequence of recent explorations by Saha et. al. expanding previous work by Bora et. al. on using Machine Learning algorithm to construct and test planetary habitability functions with exoplanet data raises important questions. The 2018 paper \citep{Potential} analyzed the elasticity of the Cobb-Douglas Habitability Score (CDHS) and compared its performance with other machine learning algorithms. They demonstrated the robustness of their methods to identify potentially habitable planets \citep{saha2018machine} from exoplanet data set. Given our little knowledge on exoplanets and habitability, these results and methods provide one important step toward automatically identifying objects of interest from large data sets by future ground and space observatories. The variable term in SBAF, $kx^{\alpha}(1-x)^{1-\alpha}$ is inspired from a history of modeling such terms as production functions and exploiting optimization principles in production economics, \citep{Saha2016}, \citep{Ginde2016}, \citep{ginde2015mining}. Complexities/bias in data may often necessitate devising classification methods to mitigate class imbalance, \citep{Mohanchandra2015} to improve upon the original method, \citep{vapnik1964}, \citep{Cortes1995} or manipulate confidence intervals \citep{Khaidem2016}. However, these improvisations led the authors to believe that, a general framework to train in forward and backward pass may turn out to be efficient. This is the primary reason to design a neural network with a novel activation function. We used the architecture to discriminate exoplanetary habitability \citep{schulze-makuch2018time}, \citep{SchulzeMakuch2011},
\citep{Irwin2014}, \citep{googlenasa}, \citep{hipparcosref}, \citep{phlref}. 
\par We had to consider the ramifications of the classification technique in astronomy. Hence, we try to classify exoplanets based on \textit{one feature of the exoplanet at a time along with multiple features of the parent star}. Note, we did not consider surface temperature, which is hard marker. An example of this is as follows:
\begin{enumerate}
    \item \textbf{Attributes of a Sample Planet}:
    \begin{itemize}
        \item Radius only
    \end{itemize}
with attributes of the Parent Star such as Mass, Radius, Effective temperature, Luminosity, Inner edge of star's habitable zone and Outer edge of star's habitable zone with
    \item \textbf{Class Attributes}:
    \begin{itemize}
        \item Thermal habitability classification label of the planet
    \end{itemize}
\end{enumerate}
Similarly, we present results of classification when we include the exoplanet’s mass, instead ofthe radius; and when we include the exoplanet’s minimum mass instead of the radius.  Reiterating, we use only one planetary attribute in a classification run.

Machine classification on habitability is a very recent area. Therefore, the motivation for contemplating such a task is beyond doubt. However, instead of using ''black-box" methods for classification, we embarked upon understanding activation functions and their role in Artificial Neural Net based classification. Theoretically, there is evidence of optima and therefore absence of local oscillations. This is significant and helps classification efficacy, for certain. In comparison to gradient boosted classification of exoplanets, \cite{Insight}, \cite{Potential}, our method achieved more accuracy, a near perfect classification. This is encouraging for future explorations into this activation function, including studying the applicability of Q-deformation and maximum entropy principles. Even without habitability classification or absence of any motivation, further study of the activation function seems promising.

Our focus has shifted from presenting and compiling the accuracy of various machine learning and data balancing methods to developing a system for classification that has practical applications and could be used in the real world. The accuracy scores that we have been able to accomplish show that with a reasonably high accuracy, the classification of the exoplanets is being done correctly. The performance of the proposed activation functions on pruned features is simply remarkable for two reasons. The first being, the pruned feature set does not contain features which account for hard markers. This makes the job hard since one would expect a rapid degradation in performance when the hard markers such as surface temperature and all its related features are removed from the feature set before classification. We note, when an exoplanet is discovered, surface temperature is one of the features Astrophysicists use to label it. If surface temperature can't be measure, it is estimated. Even if the surface temperature can't be measured, our activation functions make strong enough classifiers to predict labels of exoplanet samples, dispensing away the need for estimating it. This implies that whenever an exoplanet is newly discovered, their thermal habitability classes can be estimated using our approach. This significant information could be useful for other missions based on methods that would try following up the initial observation. Hence, samples that are interesting from a habitability point of view could gain some traction quite early.\\
Future work could focus on using adaptive learning rates \citep{ArxivRS} by fixing Lipschitz loss functions. This may help us investigate if faster convergence is achieved helping us fulfill the larger goal of parsimonious computing.
\section*{Acknowledgement}
Funding: This work was supported by the Science and Engineering Re-
search Board (SERB)-Department of Science and Technology (DST), Gov-
ernment of India (project reference number SERB-EMR/ 2016/005687). The
funding source was not involved in the study design, writing of the report,
or in the decision to submit this article for publication.


\section*{References}


\section{Appendix}
\subsection{Neural Networks with SBAF and A-ReLU}
A neural network is an interconnection of  neurons arranged in hierarchy and  predominantly used to perform predictions and classification on a dataset. Commonly, the input is given to the network over and over again to tune the network a little each time, so that when the new inputs are given, the network can predict its outcome. To explain how neural network works, we will run through a simple example of training a small network shown in figure below. Keeping its configuration simple, we have kept 2 nodes in input, hidden and output layer and have chosen SBAF and A-ReLU as activation functions to demonstrate the working of back propagation. 
\subsection{Basic Structure}
Let us assume the nodes at input layer are $i_{1}$, $i_{2}$ , at hidden layer $h_{1}$, $h_{2}$ and at output layer $o_{1}$, $o_{2}$. To start off, we assign some initial random numbers to weights and biases and move on with a forward pass demonstrated in next subsection.
\begin{figure}[htbp!]
\begin{center}
\includegraphics[scale=0.7]{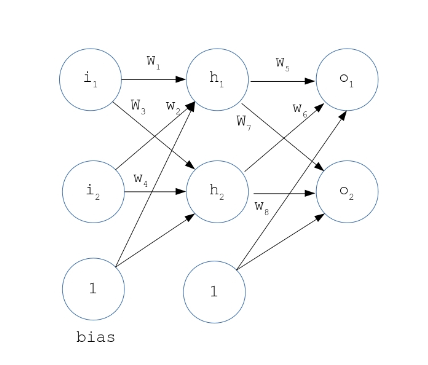} 
\caption{A simple neural network}
\label{fig:mse:1}
\end{center}
\end{figure}

\subsection{The Forward Pass using SBAF and A-ReLU}
This section computes the activation of neurons at hidden and output layers by using SBAF and A-ReLU. We start with the first entry in the data set of two inputs $i_{1}$ and $i_{2}$. The  forward pass is a linear product of inputs and weights added with a bias.
Calculating the total input at $h_1$.
$$h1_{net} = w_1 \cdot i_1 + w_2 \cdot i_2 + b_1$$
$$h2_{net} = w_3 \cdot i_1 + w_4 \cdot i_2 + b_1$$
Use SBAF to calculate the activation's of hidden neuron $h_1$ by the formula  , $y = \frac{1}{1 + kx^\alpha  (1-x)^{1-\alpha}}$.
$$h1_{out} = \frac{1}{1 + k(h1_{net})^\alpha  (1-h1_{net})^{1-\alpha}}$$
$$h2_{out} = \frac{1}{1 + k(h2_{net})^\alpha  (1-h2_{net})^{1-\alpha}}$$
In parallel, if we use A-ReLU to compute the activation's of hidden neuron $h_1$, the formula is $y = kx^n$ if $x>0$ else $y=0$, therefore
$$h1_{out} =  k(h1_{net})^n$$ 
$$h2_{out} = k(h2_{net})^n$$
assuming $h1_{net} >0$ and $h2_{net} >0$. \\
Repeat the process for neurons at output layer.
$$o1_{net} = w_5 \cdot h1_{out} + w_6 \cdot h2_{out} + b_2$$
$$o2_{net} = w_7 \cdot h1_{out} + w_8 \cdot h2_{out} + b_2$$
While using SBAF, the activation of neurons are
$$o1_{out} = \frac{1}{1 + k(o1_{net})^\alpha  (1-o1_{net})^{1-\alpha}}$$
$$o2_{out} = \frac{1}{1 + k(o2_{net})^\alpha  (1-o2_{net})^{1-\alpha}}$$
For A-ReLU, the activation are 
$$o1_{out} =  k(o1_{net})^n$$ 
$$o2_{out} = k(o2_{net})^n$$
assuming $o1_{net} >0$ and $o2_{net} >0$. \\
Since we initialized the weights and biases randomly, the outputs at the neurons are off-target. Conclusively, we need to compute the difference and propagate it back to the network. The next subsection computes the error gradient by using back propagation and adjust the weights to improve the network. Calculating the errors,
$$\mathrm{Error} = \mathrm{Error}_{o1} + \mathrm{Error}_{o2}$$
$$\mathrm{Error}_{o1} = \frac{1}{2}\left( o1_{target} - o1_{out} \right)^2$$
$$\mathrm{Error}_{o2} = \frac{1}{2}\left( o2_{target} - o2_{out} \right)^2$$
\subsection{The Backward Pass for both SBAF and A-ReLU}
This part deals with computing the error margins, the error resulted because the weights are randomly initialized. Therefore, the weights need adjustments so that the error can be decreased during predictions. Calculating the change of weights in done in two steps. The rate of change in error with respect to weights in computed in first step. In the second, the weights are updated by subtracting a portion of error gradient from weights. Similar to the forward pass, the backward pass is also computed layer-wise, but in the reverse mode. 
\subsubsection{At Output Layer}
Moving backwards, lets consider  weight $w_5$ that needs to be updated. To find the error gradient with respect to $w_5$, i.e., $\frac{\partial E_{T}}{\partial w_5}$ we use the chain rule shown in the  equation below. (Here $E_{T}$ is the total error at both neurons of output)\\
\begin{figure}[htbp!]
\begin{center}
\includegraphics[scale=0.7]{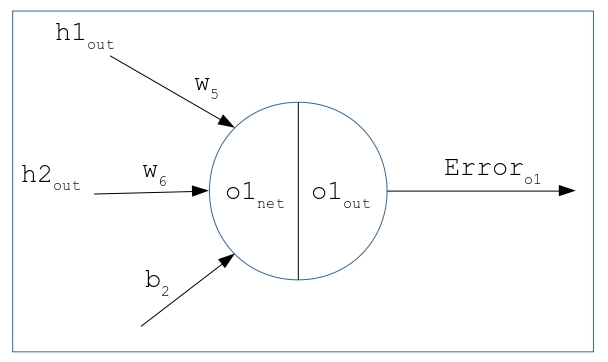} 
\caption{A simple neuron at the output layer }
\label{fig:mse:2}
\end{center}
\end{figure}

\[
\boxed{\frac{\partial E_T}{\partial w_5} = \frac{\partial E_T}{\partial o1_{out}}\cdot \frac{\partial o1_{out}}{\partial o1_{net}} \cdot \frac{\partial o1_{net}}{\partial w_5}}
\]
Taking each component one at a time on the RHS of the equation, we first derive $\frac{\partial E_T}{\partial w_5}$ -
\begin{equation}
\boxed{
	\begin{aligned} 
		 E_T &= E_{o1} + E_{o2} \\ 
		E_T &= \frac{1}{2} \left( o1_{target} - o1_{out} \right)^2 + \frac{1}{2}\left( o2_{target} - o2_{out} \right)^2 \\ 
		\frac{\partial E_T}{\partial o1_{out}} &=  2 \cdot \frac{1}{2} \left( o1_{target} - o1_{out} \right) \cdot (-1) + 0 \\
		\frac{\partial E_T}{\partial o1_{out}} &= -\left( o1_{target} - o1_{out} \right)
	\end{aligned}
} \qquad
\end{equation}
Next we compute the derivative of the activation functions in terms of $\frac{\partial o1_{out}}{\partial o1_{net}}$.  We are using SBAF and A-ReLU, and dervatives of both the functions are available. First, while using SBAF -
\begin{equation}
\boxed{
	\begin{aligned}
		o1_{out} &= \frac{1}{1 + k(o1_{net})^\alpha (1-o1_{net})^{1-\alpha}} \\
		\frac{\partial o1_{out}}{\partial o1_{net}} &= \frac{o1_{out}(1 - o1_{out})}{o1_{net}(1-o1_{net})} \cdot (\alpha - o1_{net})
	\end{aligned}
} \qquad
\end{equation}
While using A-ReLU,
\begin{equation}
\boxed{
	\begin{aligned}
		o1_{out} &= k.(o1_{net})^{n} \\
		\frac{\partial o1_{out}}{\partial o1_{net}} &= n.k^{\frac{1}{n}}.(o1_{out})^{\frac{n-1}{n}}
	\end{aligned}
} \qquad
\end{equation}
Finally the third component of the chain rule $\frac{\partial o1_{net}}{\partial w_5}$ (this is common for both SBAF and A-ReLU),
\begin{equation}
\boxed{
	\begin{aligned}
		o1_{net} &= w_5 \cdot h1_{out} + w_6 \cdot h2_{out} + b_2 \\
		\frac{\partial o1_{net}}{\partial w_5} &= h1_{out}
	\end{aligned}
} \qquad
\end{equation}
The error gradient with respect to $w_5$ for SBAF is derivable by putting $(21)$ and $(22)$ and $(24)$ together in $\frac{\partial E_T}{\partial w_5}$,
\begin{equation*}
\frac{\partial E_T}{\partial w_5} = -\left( o1_{target} - o1_{out} \right) \cdot \frac{o1_{out}(1 - o1_{out})}{o1_{net}(1-o1_{net})} \cdot (o1_{net}-\alpha) \cdot h1_{out}
\end{equation*}

Likewise the other gradients are also computed as,
\begin{equation*}
\frac{\partial E_T}{\partial w_6} = -\left( o1_{target} - o1_{out} \right) \cdot \frac{o1_{out}(1 - o1_{out})}{o1_{net}(1-o1_{net})} \cdot (o1_{net}-\alpha) \cdot h2_{out} \\
\end{equation*}
\begin{equation*}
\frac{\partial E_T}{\partial w_7} = -\left( o2_{target} - o2_{out} \right) \cdot \frac{o2_{out}(1 - o2_{out})}{o2_{net}(1-o2_{net})} \cdot (o2_{net}-\alpha) \cdot h1_{out}
\end{equation*}
\begin{equation*}
\frac{\partial E_T}{\partial w_8} = -\left( o2_{target} - o2_{out} \right) \cdot \frac{o2_{out}(1 - o2_{out})}{o2_{net}(1-o2_{net})} \cdot (o2_{net}-\alpha) \cdot h2_{out}
\end{equation*}
Correspondingly, the error gradient with respect to $w_5$ for A-ReLU is derived by keeping (21)(23) and (24) together,

\begin{equation*}
\frac{\partial E_T}{\partial w_5} = - n.k^{\frac{1}{n}} \left(o1_{target} - o1_{out} \right) \cdot (o1_{out})^{\frac{n-1}{n}} \cdot h1_{out}
\end{equation*}

Likewise the other gradients are also computed as,

\begin{equation*}
\frac{\partial E_T}{\partial w_6} = - n.k^{\frac{1}{n}} \left(o1_{target} - o1_{out} \right) \cdot (o1_{out})^{\frac{n-1}{n}} \cdot h2_{out}
\end{equation*}

\begin{equation*}
\frac{\partial E_T}{\partial w_7} = - n.k^{\frac{1}{n}} \left(o2_{target} - o2_{out} \right) \cdot (o2_{out})^{\frac{n-1}{n}} \cdot h1_{out}
\end{equation*}

\begin{equation*}
\frac{\partial E_T}{\partial w_8} = - n.k^{\frac{1}{n}} \left(o2_{target} - o2_{out} \right) \cdot (o2_{out})^{\frac{n-1}{n}} \cdot h2_{out}
\end{equation*}
Both activation functions, the weights are adjusted as -
\begin{equation}
w_5^{new} = w_5 - \eta \cdot \frac{\partial E_T}{\partial w_5}
\end{equation}
where $\eta$ is the learning rate. \\

\subsubsection{Hidden Layer} Continuing the backward pass, this part demonstrate the computation of error gradients with respect to weights that are connecting input and hidden layer. Once the gradients are found, the weights can be updated by the formula used previously. Thus, we need to derive $\frac{\partial E_T}{\partial w_1}$
 and correspondingly the other gradients can be obtained.\\
\includegraphics[scale=0.5]{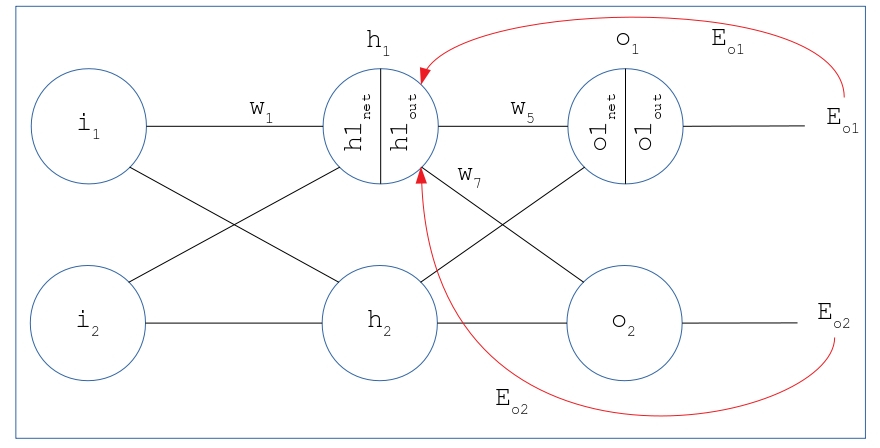}  \\
We need to find $\frac{\partial E_T}{\partial w_1}$. \\
Apparently, if $E_T$ is the total error at output layer, then \[\frac{\partial E_T}{\partial w_1} = \frac{\partial E_{o1}}{\partial w_1} + \frac{\partial E_{o2}}{\partial w_1}\]
Computing both the additive terms by using chain rule,
\setcounter{equation}{0}
\begin{equation}
\frac{\partial E_{o1}}{\partial w_1} = \frac{\partial E_{o1}}{\partial o1_{out}} \cdot \frac{\partial o1_{out}}{\partial o1_{net}} \cdot \frac{\partial o1_{net}}{\partial h1_{out}} \cdot \frac{\partial h1_{out}}{\partial h1_{net}} \cdot \frac{\partial h1_{net}}{\partial w_1}
\end{equation}
\begin{equation}
\frac{\partial E_{o2}}{\partial w_1} = \frac{\partial E_{o2}}{\partial o2_{out}} \cdot \frac{\partial o2_{out}}{\partial o2_{net}} \cdot \frac{\partial o2_{net}}{\partial h1_{out}} \cdot \frac{\partial h1_{out}}{\partial h1_{net}} \cdot \frac{\partial h1_{net}}{\partial w_1}
\end{equation}
Finding the multiplicative terms of equation $(1)$ (Please note that this is with reference to SBAF. For A-ReLU, a similar procedure is followed.),
\begin{align*}
\frac{\partial E_{o1}}{\partial o1_{out}} &= -(o1_{target} - o1_{out}) \\
\frac{\partial o1_{out}}{\partial o1_{net}} &= \frac{o1_{out}(1-o1_{out})}{o1_{net}(1-o1_{net})} \cdot (\alpha - o1_{net}) \\
\frac{\partial o1_{net}}{\partial h1_{out}} &= w_5 \left(\because o1_{net} = w_5 \cdot h1_{out} + w_6 \cdot h2_{out} + b_2 \text{ and } \frac{\partial o1_{net}}{\partial h1_{out}} = w_5 \right) \\
\frac{\partial h1_{out}}{\partial h1_{net}} &= \frac{h1_{out}(1-h1_{out})}{h1_{net}(1-h1_{net})} \cdot (\alpha - h1_{net}) \\
\frac{\partial h1_{net}}{\partial w_1} &= i_1 \left(\because h1_{net} = w_1 i_1 + w_2 i_2 + b_1 \text{ and } \frac{\partial h1_{net}}{\partial w_1} = i_1 + 0 \right)
\end{align*}
Similarly, computing all the components of $(2)$,
\begin{align*}
\frac{\partial E_{o2}}{\partial o2_{out}} &= -(o2_{target} - o2_{out}) \\
\frac{\partial o2_{out}}{\partial o2_{net}} &= \frac{o2_{out}(1-o2_{out})}{o2_{net}(1-o2_{net})} \cdot (o2_{net}-\alpha) \\
\frac{\partial o2_{net}}{\partial h1_{out}} &= w_7 \left(\because o2_{net} = w_7 h1_{out} + w_8 h2_{out} + b2 \right) \\
\end{align*}
We already know the values of $\frac{\partial h1_{out}}{\partial h1_{net}}$ and $\frac{\partial h1_{net}}{\partial w_1}$. Similar calculations hold valid for computing gradient of A-ReLU. 
\\ \\
Adding up everything,
\begin{equation*}
\frac{\partial E_T}{\partial w_1} = \frac{\partial E_{o1}}{\partial w_1} + \frac{\partial E_{o2}}{\partial w_2}
\end{equation*}
Adjusting the weight
\begin{equation*}
\boxed{
	w_1^{new} = w_1 - \eta\cdot \frac{\partial E_T}{\partial w_1}
} \qquad
\end{equation*}
Likewise, the remaining error derivatives, $\frac{\partial E_T}{\partial w_2}$, $\frac{\partial E_T}{\partial w_2}$, $\frac{\partial E_T}{\partial w_3}$, and $\frac{\partial E_T}{\partial w_4}$ can be computed in the similar manner for SBAF as well as for A-ReLU. Their corresponding weights are adjusted by using the same weight-update formula.

\newpage
\subsection*{A Mind map of key ideas presented in the paper}

\begin{figure} [!h]
  \centering
  \includegraphics[scale=0.29]{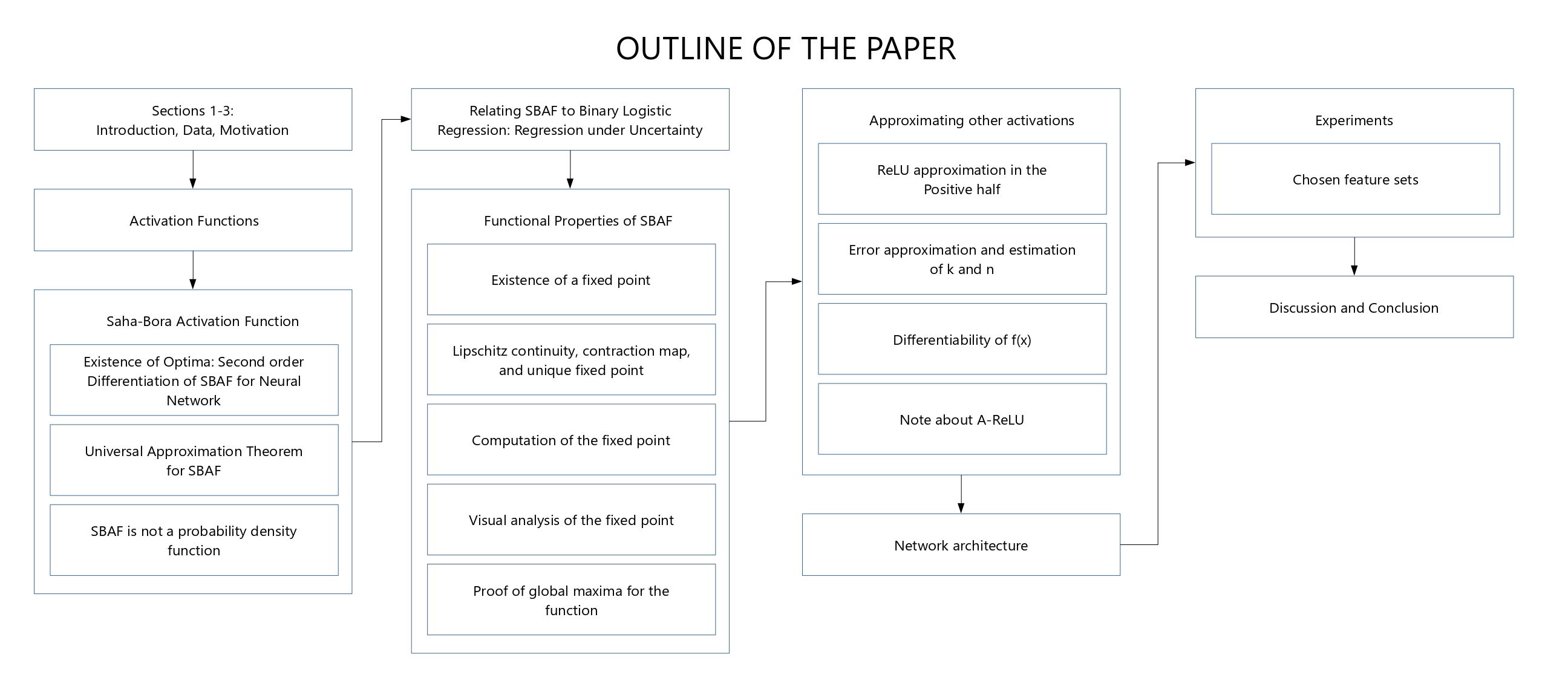}
  \caption{The flowchart is a visual description of integration of several ideas leading to novel activation functions and subsequent use in habitability classification of exoplanets.}
  \label{fig:boat1}
\end{figure}
\end{document}